\documentclass[conference,compsoc]{IEEEtran}
\ifCLASSOPTIONcompsoc
  \usepackage[nocompress]{cite}
\else
  \usepackage{cite}
\fi
\ifCLASSINFOpdf
\else
\fi
\usepackage{xurl}
\usepackage{graphicx} 
\usepackage{booktabs} 
\usepackage{algorithm}
\usepackage{algpseudocode}
\usepackage{url}
\usepackage{xcolor}
\usepackage{tabularx}
\usepackage{multirow}
\usepackage{makecell}
\usepackage{enumitem}
\usepackage{amsmath}
\usepackage{amssymb}
\usepackage[table]{xcolor}
\usepackage[most]{tcolorbox}
\definecolor{vrow}{gray}{0.90}
\definecolor{alrow}{RGB}{234,243,255}
\usepackage{subcaption}

\newtcolorbox{promptbox}[1]{
    enhanced,
    title=#1,
    width=0.98\linewidth,
    center,
    colback=white,
    colframe=gray!60,
    colbacktitle=black!85,
    coltitle=white,
    fonttitle=\bfseries\itshape,
    arc=1.5mm,
    boxrule=0.6pt,
    left=1.5mm,
    right=1.5mm,
    top=1.5mm,
    bottom=1.5mm,
    before skip=2pt,
    after skip=2pt,
    breakable
}
\newtcolorbox{promptbox_nt}{
    enhanced,
    width=0.98\linewidth,
    center,
    colback=white,
    colframe=gray!60,
    arc=1.5mm,
    boxrule=0.6pt,
    left=1.5mm,
    right=1.5mm,
    top=1.5mm,
    bottom=1.5mm,
    before skip=2pt,
    after skip=2pt,
    breakable
}
\newtcolorbox{skill_box}[1]{
    enhanced,
    width=0.98\linewidth,
    center,
    breakable,
    title={#1},
    fonttitle=\bfseries\normalsize,
    coltitle=black,
    colbacktitle=black!8,
    colback=white,
    fontupper=\color{black!75},
    boxrule=0pt,
    colframe=black,
    frame style={
        draw=black,
        line width=0.6pt,
        dash pattern=on 3pt off 3pt
    },
    arc=0pt,
    left=5pt,
    right=5pt,
    top=5pt,
    bottom=5pt,
    toptitle=7pt,
    bottomtitle=7pt,
    before skip=3pt,
    after skip=3pt
}
\newcommand{\skillupdate}[1]{%
  \par\smallskip\noindent
  \colorbox{black!6}{%
    \parbox{\dimexpr\linewidth-2\fboxsep\relax}{#1}%
  }%
  \par\smallskip
}
\definecolor{FindingBG}{HTML}{EEF2F6}      
\definecolor{FindingBorder}{HTML}{C5CFDA}  
\definecolor{FindingAccent}{HTML}{536B83}  
\definecolor{FindingText}{HTML}{354B61}    
\newtcolorbox{findingbox}{
    enhanced,
    width=\linewidth,
    colback=FindingBG,
    colframe=FindingBorder,
    boxrule=0.35pt,
    borderline west={2pt}{0pt}{FindingAccent},
    arc=1mm,
    left=7pt,
    right=7pt,
    top=5pt,
    bottom=5pt,
    boxsep=0pt,
    before skip=6pt,
    after skip=6pt
}
\begin{document}
%
\title{From Service Access to Capability Theft: Cloning Task-Solving Capabilities of Proprietary LLM Agents onto Weak Open-Weight Agents}

\author{
{\rm
Xiaoting Lyu$^{1}$\thanks{Email: \texttt{xiaoting.lyu@xjtu.edu.cn}},
Yuhong Wu$^{1}$,
Yufei Han$^{2}$,
Shichang Liu$^{1}$,
Liang Zhang$^{3}$
}\\
{\rm
Bin Wang$^{1}$,
Bin Wang$^{4}$,
Xiaobo Ma$^{1}$,
Wei Wang$^{1}$\thanks{Email: \texttt{wei.wang@xjtu.edu.cn}}
}\\[2mm]
$^{1}$Xi'an Jiaotong University
\qquad
$^{2}$INRIA
\qquad
$^{3}$University of Warwick
\\
$^{4}$Zhejiang Key Laboratory of Artificial Intelligence of Things (AIoT)
Network and Data Security
}

\maketitle

\begin{abstract}
Proprietary LLM agents provide valuable specialized task-solving capabilities, yet limited service access may expose them to unauthorized capability theft. We investigate a previously underexplored strong-to-weak capability theft threat, where an adversary seeks to reconstruct a stronger proprietary agent's task-solving capabilities on a weaker, attacker-controlled agent. Our motivating study reveals a skill execution gap: even with access to the victim's authentic skills, a weaker agent fails to reproduce its task-solving performance, highlighting the execution asymmetry between stronger and weaker agents. Our key insight is that observable victim executions expose procedural knowledge that, when adapted to the execution capabilities of weaker agents, can enable capability reconstruction. We propose \textsc{CapaTheft}, which reconstructs capability-critical behaviors through workflow abstraction, differential gap diagnosis, and structured skill rewriting, without modifying the weaker agent's model, harness, or tools. Across 20 task scenarios comprising 600 instances and eight victim-{}-attacker configurations, \textsc{CapaTheft} improves task pass rates by over 30\% compared with direct skill reuse and recovers over 80\% of the victim's capability advantage. Further evaluations demonstrate generalization to unseen instances, cross-system transferability, and the limited effectiveness of existing defenses. Our findings reveal that observable execution information, when adapted to the execution capabilities of weaker agents, can enable strong-to-weak task-solving capability theft without further victim access.
\end{abstract}


%
\IEEEpeerreviewmaketitle

\section{Introduction}
\label{sec:intro}

LLM-based autonomous agents are increasingly deployed as specialized commercial services across high-stakes domains such as software engineering, legal analysis, financial reporting, and document processing~\cite{anthropic2024agents,anthropic2025agentskills}. These systems combine foundation models, domain skills, execution harnesses, and external tools to perform multi-step planning, tool coordination, feedback interpretation, and result verification. Through the coordinated operation of these components, agents realize a \emph{reusable task-solving capability} that enables them to reliably handle complex task families across diverse inputs. This end-to-end task-solving capability constitutes a core source of commercial differentiation and competitive advantage for agent service providers, allowing them to deliver specialized services that are difficult to replicate. Building and calibrating such capabilities requires substantial domain expertise and engineering investment, making them valuable proprietary assets. For instance, Harvey, which deploys specialized agents for contract analysis, due diligence, and compliance workflows, reached an \$11 billion valuation in March 2026~\cite{harvey2026funding}. Safeguarding this competitive advantage therefore requires protecting the task-solving capabilities themselves, rather than focusing exclusively on individual components such as model weights, system prompts, or isolated skill artifacts.

Reproducing these proprietary capabilities independently requires substantial engineering effort, while continuously relying on commercial agent services incurs recurring costs and limits operational control. These incentives motivate adversaries to seek a lower-cost, independently deployable alternative by reconstructing specialized capabilities on locally hosted, weaker open-weight agents. This gives rise to a new security threat, which we refer to as \emph{strong-to-weak capability theft}: an adversary with limited access to a stronger proprietary agent seeks to reproduce its specialized task-solving capabilities on a weaker, attacker-controlled agent. For instance, an adversary with access to a proprietary commercial-contract analyzer may seek to clone its specialized risk-identification capabilities onto a lightweight open-weight agent. Doing so avoids the prohibitive cost of developing custom legal workflows from scratch or paying continuous API fees to the victim service. If successful, the attacker gains the ability to process proprietary legal documents locally at the victim's performance level without incurring further query costs, thereby eroding the victim's commercial moat.

A seemingly straightforward approach to strong-to-weak capability theft is to obtain the victim's skills, which encode reusable task procedures in portable artifacts that can be transferred across agents without accessing private model parameters. Existing studies have explored extracting or reconstructing such skills through interface probing~\cite{wang2026skillstealing}, execution trajectories~\cite{geng2026sigleak}, black-box interactions~\cite{hua2026skillclone,tsai2026daydreaming}, and execution-based skill optimization~\cite{ni2026trace2skill,yang2026skillopt,lu2026cobraskills,ma2026skillneo}. However, existing approaches focus primarily on skill recovery or optimization, overlooking the operational asymmetry of transferring a stronger agent's task-solving capability to a substantially weaker agent. In particular, skills designed for stronger agents may implicitly rely on capabilities in reasoning, tool coordination, verification, and error recovery that weaker agents struggle to realize, even when equipped with identical instructions. Thus, whether acquiring the victim's skills alone is sufficient to reproduce its task-solving capabilities on a weaker agent remains an open question.

\begin{figure}[t]
    \centering
    \includegraphics[width=\columnwidth]{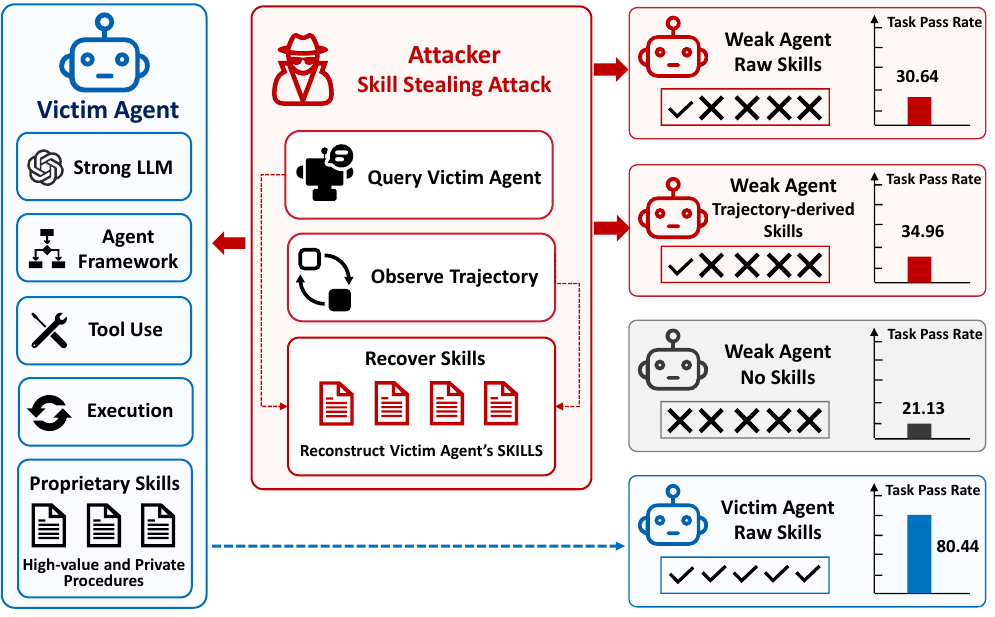}
    \caption{Task success rates under different agents.}
    \label{fig:teaser}
\end{figure}
To investigate this question, we conduct a motivating study comparing a proprietary Codex agent powered by GPT-5.5~\cite{openai2026gpt55} with a weaker OpenHands~\cite{wang2024openhands} agent powered by Qwen3.6-35B-A3B~\cite{qwen2026qwen36} across identical task scenarios. Even when equipped with the victim's raw skills, the weaker agent achieves a task pass rate of only 30.64\%, compared with 80.44\% for the victim (Figure~\ref{fig:teaser}). Replacing raw skills with those generated by Trace2Skill~\cite{ni2026trace2skill} yields only a marginal improvement to 34.96\%. We term this pronounced disparity under matched-skill conditions the \emph{skill execution gap}. This gap highlights the distinction between possessing procedural knowledge and effectively executing it: while a stronger agent can operationalize high-level skill instructions through implicit runtime decisions, such as intermediate verification and error recovery, a weaker agent may fail to reliably realize these behaviors. These results suggest that extracting or optimizing skill artifacts alone is insufficient to reproduce a stronger agent's task-solving capabilities on a weaker one. This raises a fundamental yet largely unexplored research question: \textit{\textbf{Can an adversary bridge the skill execution gap and reproduce a stronger agent's task-solving capabilities on a weaker agent while keeping its model, harness, and tools unchanged?}}

At first glance, the skill execution gap may appear to impede strong-to-weak capability theft. Our key insight, however, is that execution traces exposed through normal service interactions reveal how the victim carries out its skills, including concrete verification checks, interpretations of tool feedback, and recovery actions that are not explicitly specified in static skill artifacts. By contrasting successful victim executions with failed executions of the weaker agent, an adversary can identify capability-critical behaviors that the weaker agent fails to realize and translate them into explicit guidance adapted to its execution capabilities. However, turning such limited observations into reusable procedural guidance presents 3 technical challenges:

\noindent\textit{\textbf{Challenge 1: Instance-Specific Execution Evidence.}} Victim execution traces intertwine reusable procedural logic with task-specific actions, parameters, and intermediate artifacts. In multi-skill tasks, they may further obscure workflow-stage boundaries and dependencies across skills. Consequently, recovering generalizable task structures from a single, instance-specific execution is challenging.

\noindent\textit{\textbf{Challenge 2: Ambiguous Execution Differences.}}
Differences between successful victim executions and weaker-agent failures do not necessarily explain task failures. The two agents may follow different but equally valid tool sequences, while the weaker agent may fail to execute procedures already specified in its skills. Moreover, observed errors may be downstream consequences of earlier mistakes or stochastic variations rather than the actual causes of failure. Identifying which execution deficits are responsible for task failures is therefore challenging.

\noindent\textit{\textbf{Challenge 3: Weaker-Agent Capability Constraints.}}
Identifying a missing behavior does not guarantee that the weaker agent can execute it under fixed model and harness constraints. High-level instructions may leave critical execution decisions unresolved, whereas directly imitating the victim's actions may exceed the weaker backbone's reasoning and tool-coordination capabilities. The challenge is therefore to translate identified deficits into explicit, actionable procedures that the weaker agent can reliably execute.

 \noindent\textit{\textbf{Our Attack Approach.}} To address these challenges, we propose \textsc{CapaTheft}, a strong-to-weak capability theft approach that reconstructs a proprietary agent's task-solving capabilities on a fixed weaker agent through skill adaptation alone, keeping its model, harness, and tools unchanged. Given an initial skill collection, interface-exposed victim execution traces, and local execution feedback from the weaker, adversary-controlled agent, \textsc{CapaTheft} operates in three stages: (1)~\emph{Workflow Abstraction and Skill Mapping} addresses Challenge~1 by abstracting successful victim traces into a parameterized, stage-level workflow and mapping each stage to its relevant skills, separating reusable procedures from instance-specific details; (2)~\emph{Differential Gap Diagnosis} addresses Challenge~2 by comparing the weaker agent's failure traces against the victim's successful reference traces to identify capability-critical behavioral deficits and distinguish them from benign execution variations; and (3)~\emph{Structured Skill Rewriting} addresses Challenge~3 by translating diagnosed deficits into explicit behavioral primitives and compiling them into localized skill patches tailored to the weaker agent's capabilities. After iterative local refinement, the resulting adapted skills are frozen for independent deployment on unseen tasks without further victim access.

We evaluate \textsc{CapaTheft} across 20 task families spanning six application domains, pairing four stronger victim systems with two weaker attacker agents (eight victim--attacker configurations). For each scenario and each victim--attacker pair, using only a \emph{single} task instance to query the victim agent for capability cloning and evaluating the \textsc{CapaTheft}-curated skills across 29 held-out instances with the attacker agent, \textsc{CapaTheft} achieves the average task pass rates ($\overline{\mathrm{IPR}}$) of 70.77\%--83.72\%. It outperforms direct reuse of query-extracted skills by 35.18\%--49.52\%. Attribution experiments further show that incorporating successful victim executions increases $\overline{\mathrm{IPR}}$ from 43.28\% to 71.27\%, demonstrating that capability recovery stems from cross-agent execution contrasts rather than intrinsic model knowledge alone. The \textsc{CapaTheft}-curated skills also transfer across different model backbones and agent harnesses, while the evaluated defenses only partially mitigate capability leakage. Collectively, these findings reveal a practical strong-to-weak capability theft threat, enabling reuse of proprietary task-solving capabilities without further victim access.

Our contributions are summarized as follows:
\begin{itemize}[leftmargin=*]
    \item We identify a new {strong-to-weak capability theft} threat, showing that skill acquisition alone is insufficient for capability replication, while observable executions expose additional exploitable procedural knowledge.

    \item We propose \textsc{CapaTheft}, which reconstructs stronger agents' task-solving capabilities on weaker agents through workflow abstraction, differential diagnosis, and structured skill rewriting.

    \item We evaluate \textsc{CapaTheft} across 20 scenarios, 6 domains, 600 instances, and 8 victim--adversary configurations, demonstrating effective capability recovery, cross-system transferability, and limitations of existing defenses.
\end{itemize}

\section{Background and Related Work}
\label{sec:related_work}
\subsection{LLM Agents, Skills, Execution, and Capability}
\label{sec:background_agents}

LLM agents integrate foundation models, execution harnesses, and tools to perform multi-step tasks~\cite{anthropic2024agents,wang2024openhands}. The model interprets instructions and observations to reason about tasks and select actions~\cite{yao2023react}, while the
harness manages the execution context, interaction loop, and tool invocation~\cite{yang2024sweagent,wang2024openhands}. Skills provide reusable procedural guidance through packages of instructions, scripts, and reference materials~\cite{anthropic2025agentskills,jiang2026agenticskills}. They specify task workflows, tool-use procedures, execution constraints, and verification requirements, enabling task specialization without modifying model parameters. We refer to such explicit procedural packages as \emph{skill artifacts}.

\emph{Execution} refers to the actions through which an agent carries out task procedures, including tool invocation, feedback interpretation, error recovery, and result verification. These interactions may generate records such as tool calls, responses, and intermediate artifacts, some of which are exposed through service interfaces. For example, Codex provides terminal logs and test outputs for execution inspection~\cite{openai2025codex}. Such records reveal externally observable agent behaviors without necessarily exposing private model states or internal configurations.

We define \emph{task-solving capability} as an agent system's ability to solve instances across a task family. Unlike skill artifacts, which encode procedural guidance, task-solving capability reflects an agent's ability to effectively execute these procedures and complete tasks. It emerges from the joint operation of the model, harness, skills, tools, and execution environment. Prior studies show that agent performance depends on interface design~\cite{yang2024sweagent}, while skill utility varies across agent configurations and models~\cite{li2026skillsbench,huang2026skilllens}. Thus, identical skill artifacts do not necessarily yield equivalent task-solving capabilities across agents.

\subsection{Extraction Attacks on LLM Agents}
\label{sec:background_extraction}
Prior extraction attacks against LLM-based systems target different types of proprietary information. \emph{Model extraction} uses black-box API queries to recover model-specific information or train behavioral surrogates~\cite{carlini2024stealing,zhao2025model_extraction,liang2025lord}, while \emph{prompt extraction} targets hidden system instructions and configurations through direct disclosure or functional approximation~\cite{zhang2024effectiveprompt,tan2025promptstealing,chen2026llmthief,yang2025prsa}. More recently, \emph{skill extraction} targets reusable procedural artifacts in LLM agents by probing interfaces for skill content~\cite{wang2026skillstealing}, inferring hidden skills from execution trajectories~\cite{geng2026sigleak}, or reconstructing skill functionality through black-box interactions~\cite{hua2026skillclone,tsai2026daydreaming}.

Although these attacks recover or approximate proprietary components, they do not directly address whether a weaker agent can reproduce a stronger agent's end-to-end task-solving capabilities. In particular, recovering authentic skill artifacts does not guarantee that a weaker agent can effectively execute the encoded procedures, as demonstrated by the skill execution gap in Section~\ref{sec:intro}. In contrast, this work targets the victim's realized task-solving capability, rather than the fidelity of individual recovered artifacts. It investigates whether such capabilities can be independently reproduced on a weaker, adversary-controlled agent through limited victim access, without modifying the agent's underlying model, harness, or tools.

\subsection{Trajectory-Based Skill Optimization}
\label{sec:background-skill-optimization}

A line of research leverages execution experience to learn and refine reusable skills. Trace2Skill distills multi-trajectory lessons into transferable procedural guidance~\cite{ni2026trace2skill}, while SkillOpt~\cite{yang2026skillopt} and SkillCat~\cite{chen2026skillcat} refine skills using execution feedback and comparisons between successful and failed rollouts. COBRA-Skills further employs contextual-bandit optimization~\cite{lu2026cobraskills}, while SkillNEO explores evolutionary skill refinement~\cite{ma2026skillneo}.

Despite these advances, existing methods primarily learn or optimize reusable skills for better task performance, rather than transfer capabilities across agents with different execution capabilities. Skills distilled from stronger agents may retain implicit decisions that weaker agents cannot reliably execute, while feedback from weaker agents alone provides limited guidance for behaviors they consistently fail to perform. Moreover, existing comparisons of successful and failed executions primarily serve skill refinement rather than addressing cross-agent execution asymmetry. In contrast, this work leverages execution differences between
stronger victim agents and weaker, adversary-controlled agents to identify capability-critical behaviors and adapt procedural guidance to weaker agents.


\section{Problem Formulation and Threat Model}
\label{sec:threat_model}
\subsection{Capability Theft Scenario and Formulation}
\label{subsec:theat_ecosystem}


We consider a strong-to-weak capability theft scenario in which an adversary accesses a proprietary agent service as a legitimate customer while operating a local agent powered by a substantially weaker open-weight model. The victim provides specialized task-solving capabilities through a black-box service interface that accepts task requests and exposes execution information, such as tool invocations, intermediate artifacts, and verification results, for user transparency and auditability. With limited access to the victim, the adversary seeks to leverage this observable execution evidence to reconstruct the victim's reusable task-solving capabilities on its weaker agent. The ultimate goal is to enable the weaker agent to independently solve unseen task instances without upgrading its underlying model or further accessing the victim.

\noindent\textbf{Agent Capability.}
We model an agent system as $A=(M,H,T,E)$, where $M$ denotes the LLM backbone, $H$ the execution harness, $T$ the available tools, and $E$ the execution environment. Let $\mathcal{D}$ denote a target task distribution and $\mathcal{K}$ a collection of skills supporting the corresponding task family. Given a task instance $x\sim\mathcal{D}$, let $\tau_{A,\mathcal{K}}(x;\omega)$ denote the execution trajectory of agent $A$ equipped with $\mathcal{K}$, where $\omega$ captures execution stochasticity. We define its task-solving capability as the expected task success rate over $\mathcal{D}$:
\begin{equation}
C_{\mathcal{D}}(A,\mathcal{K})
=
\mathbb{E}_{x\sim\mathcal{D},\,\omega}
\left[
V_x\left(\tau_{A,\mathcal{K}}(x;\omega)\right)
\right],
\label{eq:capability}
\end{equation}
where $V_x(\tau)\in\{0,1\}$ is a task-specific verifier indicating whether the execution successfully solves the  instance $x$.

\noindent\textbf{Victim--Adversary Capability Gap.}
Let $A_v$ denote the stronger proprietary victim agent equipped with its underlying skill collection $\mathcal{K}_v$, and $A_a$ the weaker, adversary-controlled agent initialized with $\mathcal{K}_{\rm init}$. The initial skills may be obtained through interface-based skill extraction~\cite{wang2026skillstealing}, trajectory-based skill reconstruction~\cite{geng2026sigleak}, or publicly available resources, and need not match the victim's actual skills. We quantify the initial victim--adversary capability gap as
\begin{equation}
\Delta_{\rm init}
=
C_{\mathcal{D}}(A_v,\mathcal{K}_v)
-
C_{\mathcal{D}}(A_a,\mathcal{K}_{\rm init})
>0.
\label{eq:initial_gap}
\end{equation}
This gap may arise from differences in both skill content and agent execution capabilities. Importantly, even when both agents are equipped with identical skills ($\mathcal{K}_{\rm init}=\mathcal{K}_v$), the weaker agent may still fail to realize behaviors that the stronger agent can reliably execute. We refer to this residual disparity under matched-skill conditions as the \emph{skill execution gap}, defined as
$\Delta_{\rm exec}=C_{\mathcal{D}}(A_v,\mathcal{K}_v)-C_{\mathcal{D}}(A_a,\mathcal{K}_v)$.
The skill execution gap captures the execution asymmetry between stronger and weaker agents, demonstrating why acquiring identical skill artifacts alone may be insufficient to reproduce a stronger agent's task-solving capabilities.

\subsection{Threat Model}
\label{sec:theat_attacker}

\noindent\textbf{Adversary Objective.}
The adversary aims to reproduce the victim's specialized task-solving capabilities by constructing an adapted skill collection $\widehat{\mathcal{K}}$ that enables the weaker agent $A_a$ to approach the performance of the victim $A_v$ equipped with its authentic skills $\mathcal{K}_v$. Starting from an initial skill collection $\mathcal{K}_{\rm init}$, the adversary seeks to minimize the skill execution gap through skill adaptation, without modifying the weaker agent's model, harness, tools, or execution environment. Formally, given a victim-query budget $B_v$, the objective is
\begin{equation}
\begin{aligned}
\min_{\widehat{\mathcal{K}}}\quad
& C_{\mathcal{D}}(A_v,\mathcal{K}_v)
- C_{\mathcal{D}}(A_a,\widehat{\mathcal{K}})\\
\mathrm{s.t.}\quad
& N_v \leq B_v,\,\,
(M_a,H_a,T_a,E_a)\ \text{remain fixed},
\end{aligned}
\label{eq:capability_theft_objective}
\end{equation}
where $N_v$ denotes the number of victim queries used during capability reconstruction, and $B_v$ specifies the corresponding query budget. For each task scenario, the adversary selects only one task instance and queries the victim on this same instance to obtain a successful reference trajectory. All query instances issued to obtain this trajectory count toward $N_v$. 
Unlike model extraction~\cite{carlini2024stealing,zhao2025model_extraction,liang2025lord}, prompt extraction~\cite{tan2025promptstealing,yang2025prsa}, and skill extraction~\cite{wang2026skillstealing,hua2026skillclone}, our attack targets the victim's realized task-solving capability, aiming to reproduce it on a substantially weaker agent rather than recover the underlying artifacts of the authentic skill.

\noindent\textbf{Adversary Capabilities.}
The adversary has limited black-box query access to the victim agent $A_v$ through its normal service interface. It can submit task instances and collect interface-exposed execution evidence, such as tool invocations, intermediate outputs, execution feedback, and final deliverables. The adversary may obtain initial skills through existing extraction techniques~\cite{wang2026skillstealing,geng2026sigleak}, interface interactions, or publicly available resources. However, it has no privileged access to the victim's internal components and cannot modify or interfere with their execution. On the adversary side, the adversary controls $A_a$ and can perform repeated local executions, analyze failures, and modify its skill collection, while keeping the underlying model, harness, tools, and environment fixed. The adversary may also employ an auxiliary LLM $M_{\rm aux}$ to analyze execution evidence and assist with skill reconstruction. $M_{\rm aux}$ is used only during curating and adapting the attacker's skills and does not execute target tasks on behalf of the weaker agent during deployment.


\noindent\textbf{Adversary Knowledge.}
The adversary knows the target task family, the specifications and inputs of its selected query instances, and its initial skill collection $\mathcal{K}_{\rm init}$. However, it does not know the complete task distribution $\mathcal{D}$ or have guaranteed access to the victim's authentic skills $\mathcal{K}_v$. Its observations are restricted to information exposed through the normal service interface, without access to the victim's model parameters, hidden system prompts, private memory, internal reasoning states, or unexposed execution records.


This threat model captures a practical strong-to-weak capability theft risk in agent services. An adversary with ordinary customer access can reproduce a proprietary agent's specialized task-solving capabilities on a substantially weaker, locally controlled agent, without compromising the victim's internal components or developing a comparably capable system. Once reconstructed, these capabilities can be independently reused on unseen tasks, allowing the adversary to circumvent the substantial costs of developing comparable capabilities or continuously relying on the victim's service. This poses a direct threat to the provider's proprietary assets, engineering investments, and competitive advantage.

\section{Preliminary Study of Capability Theft}
\label{sec:motivation}

\noindent\textbf{Experimental Setup.}
We use OpenHands~\cite{wang2024openhands} powered by Qwen3.6-35B-A3B~\cite{qwen2026qwen36} as the weaker agent and consider four victim configurations: Claude Code~\cite{anthropic2025claudecode} powered by Opus 4.8~\cite{anthropic2026opus48}, Codex~\cite{openai2025codex} powered by GPT-5.5, and OpenHands powered by DeepSeek-V4-Flash~\cite{deepseekai2026deepseekv4} or MiniMax M3~\cite{minimax2026m3}. We evaluate the weaker agent under different skill acquisition, extraction, and optimization conditions against each victim configuration, while keeping the weaker agent's model, harness, and tools fixed. Unless otherwise specified, Table~\ref{tab:motivating_results} reports average task pass rates ($\overline{\mathrm{IPR}}$s) across the four victim configurations for both the victim and the weaker agent under each skill condition. Task completion is assessed using task-specific verifiers.


\begin{table}[t]
\centering
\small
\caption{Average task pass rates across four victim configurations.}
\label{tab:motivating_results}
\small
\begin{tabular}{lll}
\toprule
Agent & Skill Condition & $\overline{\mathrm{IPR}}$ (\%) \\
\midrule
Victim $A_v$ & Authentic Skill & 76.10 \\
Victim $A_v$ & Query-Extracted Skill\cite{wang2026skillstealing} & 76.10 \\
Weaker $A_a$ & No Skill & 21.13 \\
\midrule
\multicolumn{3}{l}{\textit{Skill Acquisition and Extraction}} \\
Weaker $A_a$ & Authentic Skill & 30.64 \\
Weaker $A_a$ & Query-Extracted Skill\cite{wang2026skillstealing} & 30.64 \\
Weaker $A_a$ & SigLeak~\cite{geng2026sigleak} & 31.64 \\
\midrule
\multicolumn{3}{l}{\textit{Skill Learning and Optimization}} \\
Weaker $A_a$ & Trace2Skill~\cite{ni2026trace2skill} & 34.96 \\
Weaker $A_a$ & SkillOpt~\cite{yang2026skillopt} & 28.23 \\
\bottomrule
\end{tabular}
\end{table}

\noindent\textit{\textbf{Observation 1: Direct queries can recover near-authentic skill content.}}
Directly querying the victim through its normal interface can reveal skill content that closely matches its authentic instructions, rather than merely describing their functionality. Across the four victim configurations, the victims achieve the same $\overline{\mathrm{IPR}}$ of 76.10\% with either authentic or query-extracted skills, suggesting that the extracted content preserves the original skills' task-level functionality on the evaluated tasks. Although this does not guarantee complete recovery of all underlying skill files, it demonstrates that functionally comparable skill content can be obtained without privileged access, consistent with prior findings on black-box skill extraction~\cite{wang2026skillstealing}.


\noindent\textit{\textbf{Observation 2: Skill acquisition alone cannot reproduce stronger-agent capabilities.}}
Across the four victim configurations, the weaker agent achieves an $\overline{\mathrm{IPR}}$ of 21.13\% without skills and 30.64\% when equipped with either authentic or query-extracted skills, whereas the victim's $\overline{\mathrm{IPR}}$ with authentic skills is 76.10\%. Importantly, a substantial 45.46-percentage-point gap persists even under matched-skill conditions. We term this disparity the \emph{skill execution gap}, demonstrating that identical procedural instructions do not guarantee equivalent task-solving capabilities. While stronger agents can interpret high-level instructions, coordinate tools, and verify intermediate results, weaker agents may fail to reliably realize these behaviors.


\noindent\textit{\textbf{Observation 3: Existing skill extraction and optimization provide limited capability recovery.}}
We further evaluate whether existing skill reconstruction and optimization methods can bridge the skill execution gap. As shown in Table~\ref{tab:motivating_results}, Trace2Skill achieves the highest $\overline{\mathrm{IPR}}$ of 34.96\%, while SkillOpt achieves only 28.23\%. SigLeak also yields limited improvement, reaching 31.64\%. Even the best-performing baseline remains 41.14 percentage points below the average victim performance. These results suggest that acquiring, distilling, or optimizing skill artifacts alone is insufficient to reproduce stronger-agent capabilities on a weaker agent.


\noindent\textit{\textbf{Key Security Insight.}}
These observations reveal that acquiring or optimizing skill artifacts alone is insufficient to reproduce a stronger agent's task-solving capabilities on a weaker agent. However, observable victim executions expose additional procedural knowledge, including concrete verification checks, feedback interpretation, and error-recovery behaviors that may remain implicit in static skills. By contrasting successful victim executions with weaker-agent failures, an adversary can identify capability-critical behavioral deficits and adapt the exposed execution knowledge to the weaker agent's capabilities. This creates an opportunity to bridge the execution gap and enables strong-to-weak capability theft.

\begin{figure*}[t]
\centering
\includegraphics[width=\textwidth]{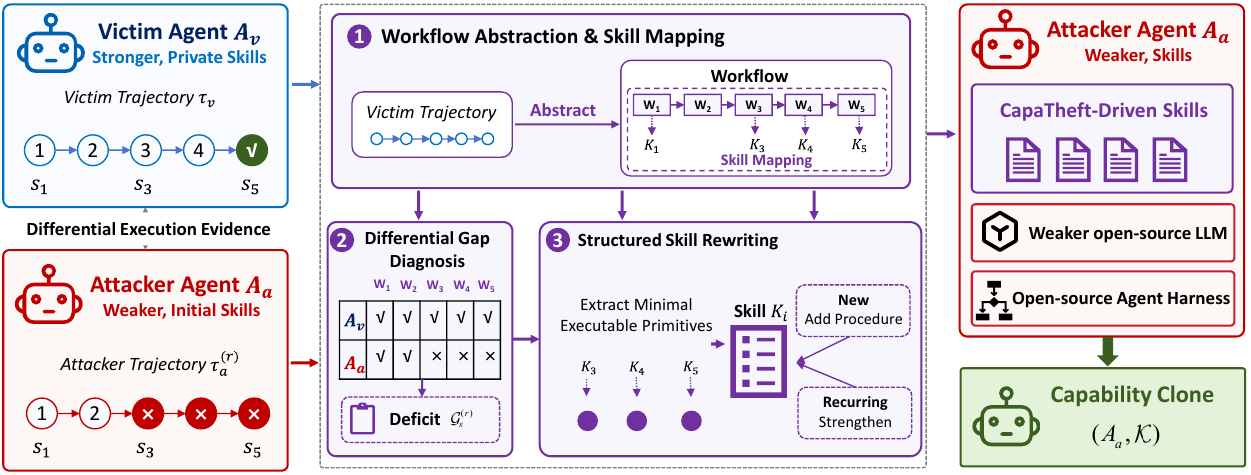}
\caption{Overview of \textsc{CapaTheft}.}
\label{fig:attack-overview}
\end{figure*}

\section{Design of CapaTheft }
\label{sec:capaclone}
\subsection{Overview and Attack Objective}
\label{subsec:capatheft_overview}

Given the capability theft objective in Section~\ref{sec:theat_attacker}, \textsc{CapaTheft} adapts an initial skill collection $\mathcal{K}_{\rm init}$, obtained through black-box queries to the victim's service interface\cite{wang2026skillstealing}, to reproduce the victim's task-solving capabilities on a weaker agent while keeping its underlying components unchanged. Since the adversary does not know the complete task distribution $\mathcal{D}$ and cannot directly evaluate its expected performance during offline reconstruction, \textsc{CapaTheft} formulates an empirical reconstruction objective over a small set of task instances $\mathcal{X}_{\rm rec}=\{x_i\}_{i=1}^{N}$. We select one reconstruction instance for each task scenario, such that the weaker agent $A_a$ initially fails to complete it with $\mathcal{K}_{\rm init}$. For each selected instance $x_i$, the adversary repeatedly queries the victim $A_v$ on the same instance until obtaining a successful reference trajectory or exhausting the query budget. Based on these reconstruction instances, we formulate an empirical objective to minimize the task-success gap between the victim and the weaker agent:
\begin{equation}
\small
\begin{aligned}
\widehat{\mathcal{K}}
\in
\underset{\mathcal{K}}{\operatorname{argmin}}
 &
\frac{1}{|\mathcal{X}_{\rm rec}|}
\sum_{x_i\in\mathcal{X}_{\rm rec}}
\mathbb{E}_{\omega_a}
\left[
r_v^{(i)}
-
V_{x_i}\left(
\tau_{A_a,\mathcal{K}}(x_i;\omega_a)
\right)
\right]
\\
\mathrm{s.t.}\quad &
N_v\leq B_v,\,\,
(M_a,H_a,T_a,E_a)\ \text{remain fixed},
\end{aligned}
\label{eq:empirical_contrastive_obj}
\end{equation}
where $\mathcal{K}$ denotes a candidate skill collection obtained by adapting $\mathcal{K}_{\rm init}$, $r_v^{(i)}=1$ indicates that the victim successfully completes $x_i$, and $B_v$ denotes the victim-query budget during capability reconstruction. The objective minimizes the empirical task-success gap between the victim and the weaker agent on the reconstruction instances. Rather than using these instances to approximate the entire task distribution, \textsc{CapaTheft} treats them as behavioral anchors for identifying and correcting capability-critical execution deficits. The resulting skill collection $\widehat{\mathcal{K}}$ is subsequently frozen and evaluated on held-out instances.

Figure~\ref{fig:attack-overview} illustrates the overall architecture of \textsc{CapaTheft}. Given a reconstruction instance $x_{\rm rec}$ and its successful victim reference trajectory $\tau_v$, \textsc{CapaTheft} performs three stages in an iterative local refinement process, each addressing one of the challenges identified in Section~1. First, \emph{Workflow Abstraction and Skill Mapping} addresses Challenge~1 by extracting reusable task structures from $\tau_v$ and associating workflow stages with relevant skills. Second, \emph{Differential Gap Diagnosis} addresses Challenge~2 by comparing the weaker agent's local execution trajectories against the victim's reference trajectory to identify capability-critical behavioral deficits. Third, \emph{Structured Skill Rewriting} addresses Challenge~3 by translating identified deficits into explicit procedural guidance adapted to the weaker agent's execution capabilities. After each skill update, the weaker agent performs new local executions to validate the modifications and identify remaining deficits. This process continues until no further capability-critical deficits are identified or the maximum refinement rounds $R_{\max}$ are reached. Throughout this refinement process, the successful victim trajectory is reused as a fixed reference, requiring no additional victim queries. The resulting skill collection $\widehat{\mathcal{K}}$ is frozen for independent deployment on unseen tasks. Section~5.2 details the three stages, and Algorithm~1 summarizes the complete reconstruction procedure.

\subsection{Technical Pipeline of CapaTheft}
\label{subsec:capability_reconstruction}
We detail the three stages of \textsc{CapaTheft} for each reconstruction instance $x_{\rm rec}\in\mathcal{X}_{\rm rec}$, using the cached successful victim trajectory $\tau_v$ as a reference. Let $\mathcal{K}^{(r)}$ denote the weaker agent's skill collection at local refinement round $r$. To account for execution stochasticity, we collect $q$ local trajectories, denoted by $\mathcal{T}_a^{(r)}=\{\tau_a^{(r,k)}\}_{k=1}^{q}$, where $\tau_a^{(r,k)}$ represents the $k$-th execution trajectory. We also maintain a history $\mathcal{H}^{(r)}$ of previously diagnosed capability deficits and applied skill patches, initialized as $\mathcal{H}^{(0)}=\varnothing$. The following subsections detail the three stages.

\subsubsection{Workflow Abstraction and Skill Mapping}
\label{subsec:workflow_abstraction}
A successful victim execution interleaves reusable procedural behaviors with instance-specific actions and intermediate artifacts. In multi-skill tasks, the execution trace may further obscure workflow-stage boundaries and implicit dependencies across skills. To address Challenge~1 in Section \ref{sec:intro}, \textsc{CapaTheft} abstracts the successful victim trajectory $\tau_v$ into a structured workflow that captures reusable task procedures and cross-skill dependencies, while mapping each workflow stage to relevant skills in $\mathcal{K}_{\rm init}$. Specifically, the auxiliary model $M_{\rm aux}$ constructs a workflow skill $S_{\rm wf}$ as follows:
\begin{equation}
\small
\begin{aligned}
S_{\rm wf}
&=
\operatorname{WorkflowExtract}\!\left(
M_{\rm aux},x_{\rm rec},\tau_v,\mathcal{K}_{\rm init}
\right)\\
&\triangleq
\left(W,Q,\Phi,R^{(0)}\right),
\end{aligned}
\label{eq:workflow_extraction}
\end{equation}
where $W$ specifies the ordered workflow stages and their procedural dependencies; $Q$ defines stage-level and final verification criteria; $\Phi$ maps each workflow stage to its relevant skills in $\mathcal{K}_{\rm init}$; and $R^{(0)}=\varnothing$ initializes an empty refinement-guidance section. When an initial skill does not cover a particular workflow stage, $S_{\rm wf}$ itself provides the corresponding procedural guidance.

More specifically, $M_{\rm aux}$ organizes the victim's execution trajectory into ordered workflow stages by grouping tool operations according to their procedural purposes and identifying dependencies between consecutive stages. Instance-specific details, such as file paths, entity names, and literal values, are replaced with parameterized references to task inputs, while reusable procedures, tool-use patterns, verification checks, and error-recovery steps are retained. For each stage $w\in W$, $M_{\rm aux}$ derives corresponding verification criteria $Q$, specifying the expected intermediate states and completion conditions for validating the stage's execution. It also identifies relevant skills based on observed tool executions and skill descriptions, establishing the mapping $\Phi(w)$. This mapping supports many-to-many relationships: a workflow stage may involve multiple skills, and a single skill may contribute to multiple stages, explicitly capturing cross-skill dependencies.
The resulting workflow skill is structured as follows:
\begin{promptbox}{Workflow Skill $S_{\rm wf}$}
\small
\noindent\texttt{Execution Structure $W$:}
Ordered execution steps and dependencies.

\noindent\texttt{Verification $Q$:}
Expected intermediate states and completion conditions.

\noindent\texttt{Skill Mapping $\Phi$:}
Mapping from workflow steps to the existing skills.

\noindent\texttt{Refinement Guidance $R^{(r)}$:}
Adaptive guidance updated with identified capability deficits.
\end{promptbox}

Before iterative refinement, $S_{\rm wf}$ is added to the initial skill collection to form
$\mathcal{K}^{(0)}=\mathcal{K}_{\rm init}\cup\{S_{\rm wf}\}$.
The structural components $(W,Q,\Phi)$ remain fixed throughout subsequent refinement rounds, providing a stable reference for differential diagnosis. Workflow-level updates are restricted to $R^{(r)}$.

\subsubsection{Differential Gap Diagnosis}
\label{subsec:gap_diagnosis}
Execution differences between the victim and weaker attack agent do not necessarily indicate capability-critical deficiencies. Different tool choices and execution paths may produce equally valid results, while behaviors already specified in the acquired skills may still be omitted or incorrectly executed by the weaker attack agent. Moreover, observed errors may be downstream consequences of earlier mistakes rather than the actual causes of task failure. To address Challenge~2 in Section~\ref{sec:intro}, \textsc{CapaTheft} performs workflow-guided differential diagnosis to identify missing or misexecuted behaviors that account for the weaker agent's failures.

At refinement round $r$, $M_{\rm aux}$ compares the cached victim trajectory $\tau_v$ with each local trajectory $\tau_a^{(r,k)}$ of the weaker agent, conditioned on its current skill collection $\mathcal{K}^{(r)}$ and diagnosis history $\mathcal{H}^{(r)}$. For each trajectory, the diagnosis produces a set of candidate capability deficits:
\begin{equation}
\small
\mathcal{G}^{(r,k)}
=
\operatorname{Diagnose}\!\left(
M_{\rm aux},
x_{\rm rec},
\tau_v,
\tau_a^{(r,k)},
\mathcal{K}^{(r)},
\mathcal{H}^{(r)}
\right).
\label{eq:gap_diagnosis}
\end{equation}
For each workflow stage $w\in W$, $M_{\rm aux}$ aligns the victim's reference behavior with the corresponding weaker-agent execution to identify missing or misexecuted procedures. The diagnosis filters out instance-specific differences and functionally valid alternative execution paths, while examining the current skill instructions and upstream execution dependencies to identify behaviors associated with task failure.

To mitigate execution stochasticity, \textsc{CapaTheft} aggregates semantically equivalent diagnoses across the $q$ local trajectories into a consolidated defect set $\mathcal{G}^{(r)}$. The occurrence frequency of each defect $g\in\mathcal{G}^{(r)}$ is calculated as:
\begin{equation}
\small
\operatorname{count}_r(g)
=
\sum_{k=1}^{q}
\mathbb{I}\!\left[
g\in\mathcal{G}^{(r,k)}
\right].
\label{eq:defect_count}
\end{equation}
Defects are prioritized in descending order of $\operatorname{count}_r(g)$, giving higher priority to persistent failure patterns. Furthermore, by consulting $\mathcal{H}^{(r)}$, $M_{\rm aux}$ distinguishes between \emph{newly identified} defects, which have not been previously addressed, and \emph{recurring} defects, which remain unresolved despite earlier skill patches. Recurring defects receive particular attention because they may require more explicit and fine-grained procedural guidance. Each diagnosed defect is represented by the following structured record:
\begin{promptbox}{Capability Defect Diagnosis $g\in\mathcal{G}^{(r)}$}
\small
\noindent\texttt{<Deficit\_id>}:
Unique identifier of the deficit.

\noindent\texttt{<Count>}:
Occurrence frequency of the deficit.

\noindent\texttt{<Target\_skill>}:
Skill responsible for the identified deficit.

\noindent\texttt{<Observation>}:
Observable failure symptom from execution trajectories.

\noindent\texttt{<Root\_reason>}:
Underlying cause of the failure.

\noindent\texttt{<Reconstruction\_status>}:
Whether the deficit is newly identified or recurring.
\end{promptbox}

\subsubsection{Structured Skill Rewriting}
\label{subsec:skill_rewriting}
Identifying capability-critical deficits does not guarantee that the weaker attack agent can execute the missing behaviors. High-level instructions may leave critical execution decisions implicit, while directly imitating the victim's actions may exceed the weaker agent's reasoning and tool-coordination capabilities. To address Challenge~3 in Section~\ref{sec:intro}, \textsc{CapaTheft} adapts the diagnosed behaviors to the weaker agent by decomposing implicit execution decisions into explicit, fine-grained procedures. Rather than modifying the underlying model, it adjusts the specificity and granularity of skill instructions to reduce reliance on implicit reasoning and complex execution decisions.

For each skill $S_j^{(r)}\in\mathcal{K}^{(r)}$ associated with diagnosed deficits $\mathcal{G}^{(r)}[S_j^{(r)}]$, the auxiliary model $M_{\rm aux}$ extracts behavioral primitives $\mathcal{P}_j^{(r)}$ tailored to the execution capabilities of the weaker agent:
\begin{equation}
\small
\mathcal{P}_j^{(r)}
=
\operatorname{ExtractPrimitive}\!\left(
M_{\rm aux},
M_a,
\mathcal{G}^{(r)}[S_j^{(r)}],
S_j^{(r)}
\right).
\label{eq:primitive_extraction}
\end{equation}
Each primitive translates a diagnosed deficit into actionable guidance by specifying the triggering conditions, concrete operations, and corresponding verification and recovery procedures, as illustrated below.
\begin{promptbox}{Behavioral Primitive $p\in\mathcal{P}_j^{(r)}$}
\small
\noindent\texttt{<Target\_skill>}:
The Skill to be refined.

\noindent\texttt{<Missing\_behavior>}:
The capability-critical behavior missing or insufficient in the attacker agent.

\noindent\texttt{<Required\_procedure>}:
The procedural guidance describing how the skill should realize missing behavior.

\noindent\texttt{<Expected\_effect>}:
The expected improvement after incorporating the primitive.
\end{promptbox}

Given the extracted primitives, $M_{\rm aux}$ incorporates them into the existing skill structure through localized patches:
\begin{equation}
\small
\Pi_j^{(r)}
=
\operatorname{GeneratePatch}\!\left(
M_{\rm aux},
\mathcal{P}_j^{(r)},
S_j^{(r)}
\right).
\label{eq:patch_generation}
\end{equation}
Each patch specifies a target location and an editing operation. Specifically, \textsc{Insert} adds missing procedural rules, whereas \textsc{Replace} decomposes or clarifies existing instructions that are too abstract, ambiguous, or repeatedly misexecuted by the weaker agent. For recurring defects identified through $\mathcal{H}^{(r)}$, the corresponding instructions are further decomposed into more explicit and granular procedures. When modifying the workflow skill $S_{\rm wf}$, patches are restricted to its refinement-guidance component $R^{(r)}$, preserving the structural components $(W,Q,\Phi)$.
\begin{promptbox}{Skill Patch $\pi\in\Pi_j^{(r)}$}
\small
\noindent\texttt{<Patch\_id>}:
Unique identifier of the patch.

\noindent\texttt{<Target\_skill>}:
Skill to be modified.

\noindent\texttt{<Location>}:
Target section or position for modification.

\noindent\texttt{<Operation>}:
Modification operation: \textsc{Insert} or \textsc{Replace}.

\noindent\texttt{<Target\_instruction>}:
The existing instruction to be replaced when the operation is \textsc{Replace}.

\noindent\texttt{<Content>}:
The new procedural instruction to be inserted or used for replacement.
\end{promptbox}

Patches across all targeted skills are aggregated into a round-level patch set $\Pi^{(r)}=\bigcup_j\Pi_j^{(r)}$, which updates the skill collection while preserving unrelated instructions:
\begin{equation}
\small
\mathcal{K}^{(r+1)}
=
\operatorname{ApplyPatch}\!\left(
\mathcal{K}^{(r)},\Pi^{(r)}
\right).
\label{eq:skill_update}
\end{equation}
The updated skills are evaluated through subsequent local rollouts to assess whether the diagnosed deficits have been corrected. Finally, $(\mathcal{G}^{(r)},\Pi^{(r)})$ is appended to $\mathcal{H}^{(r+1)}$ to guide subsequent refinement. 

\subsubsection{End-to-End Attack Procedure}
\label{subsec:attack-procedure}

Algorithm~\ref{alg:capatheft} summarizes the end-to-end reconstruction procedure of \textsc{CapaTheft}. For each task scenario, the adversary uses a single reconstruction instance and first checks whether the weaker agent can solve it with its initial skills. If not, the victim is repeatedly queried on the same instance until a successful reference trajectory is obtained or the query budget is exhausted. Using the cached trajectory, \textsc{CapaTheft} performs workflow abstraction, differential diagnosis, and structured skill rewriting in an iterative local refinement loop. Each round evaluates the updated skills through local rollouts, and the process terminates when no further deficits are identified or the maximum refinement rounds $R_{\max}$ are reached. The reconstructed skills are then frozen for independent deployment. All victim queries are counted toward the cumulative query counter $b$, while subsequent refinement requires no further victim access. A representative reconstruction example is provided in Appendix~\ref{app:newskill}.

\begin{algorithm}[t]
\caption{\textsc{CapaTheft}}
\label{alg:capatheft}
\small
\begin{algorithmic}[1]
\Require Victim $A_v$, weaker agent $A_a$, auxiliary model $M_{\rm aux}$, initial skills $\mathcal{K}_{\rm init}$, instance $x_{\rm rec}$, query budget $B_v$, current query count $b$, refinement limit $R_{\max}$, rollout count $q$
\Ensure Reconstructed skills $\widehat{\mathcal{K}}$, updated query count $b$

\State $\mathcal{K}\gets\mathcal{K}_{\rm init}$; $\mathcal{H}\gets\varnothing$

\State $(\tau_a,\mathit{ok}_a)\gets\operatorname{Exec}(A_a,\mathcal{K},x_{\rm rec})$
\If{$\mathit{ok}_a=1$}
    \State \Return $(\mathcal{K},b)$
\EndIf

\State $\mathit{ok}_v\gets 0$
\While{$\mathit{ok}_v=0$ \textbf{and} $b<B_v$}
    \State $(\tau_v,\mathit{ok}_v)\gets\operatorname{Query}(A_v,x_{\rm rec})$
    \State $b\gets b+1$
\EndWhile
\If{$\mathit{ok}_v=0$}
    \State \Return $(\bot,b)$
\EndIf
\State $S_{\rm wf}\gets\operatorname{WorkflowExtract}(M_{\rm aux},x_{\rm rec},\tau_v,\mathcal{K})$
\State $\mathcal{K}\gets\mathcal{K}\cup\{S_{\rm wf}\}$
\State $\mathcal{T}_a\gets\operatorname{Exec}^{q}(A_a,\mathcal{K},x_{\rm rec})$

\For{$r=1$ to $R_{\max}$}
    \State $\mathcal{G}\gets\operatorname{Diagnose}(M_{\rm aux},x_{\rm rec},\tau_v,\mathcal{T}_a,\mathcal{K},\mathcal{H})$
    \If{$\mathcal{G}=\varnothing$}
    \State \textbf{break}
\EndIf
    \State $\Pi\gets\operatorname{StructuredRewrite}(M_{\rm aux},\mathcal{G},\mathcal{K})$
    \State $\mathcal{K}\gets\operatorname{ApplyPatch}(\mathcal{K},\Pi)$
    \State $\mathcal{H}\gets\mathcal{H}\cup\{(\mathcal{G},\Pi)\}$
    \State $\mathcal{T}_a\gets\operatorname{Exec}^{q}(A_a,\mathcal{K},x_{\rm rec})$
\EndFor

\State $\widehat{\mathcal{K}}\gets\mathcal{K}$
\State \Return $(\widehat{\mathcal{K}},b)$
\end{algorithmic}
\end{algorithm}

\section{Empirical Evaluation}
\label{sec:evaluation}
\subsection{Experimental Setup}
\label{subsec:setup}

\subsubsection{Datasets and Task Instances}
\label{subsec:datasets}
We evaluate \textsc{CapaTheft} on \textsc{SkillsBench}~\cite{li2026skillsbench}, a benchmark of realistic agentic tasks involving multi-step planning, tool interactions, specialized skills, and task-specific verification. To study capability theft beyond skill acquisition, we select 20 task scenarios exhibiting a clear \emph{skill execution gap} under matched-skill conditions: the victim successfully completes the tasks using its authentic skills, whereas the weaker agent fails even when equipped with the same skills. The selected scenarios span six domains: office document processing and form automation, software development and debugging, financial analysis and reporting, scientific data analysis, industrial control and equipment maintenance, and formal planning. They cover diverse workflows and skill compositions, with difficulty levels ranging from easy to hard. Table~\ref{tab:selected-tasks} summarizes the selected scenarios.

\begin{table}[t]
    \centering
    \caption{Task scenarios used in our evaluation.}
    \label{tab:selected-tasks}
    \scriptsize
    \setlength{\tabcolsep}{2.2pt}
    \renewcommand{\arraystretch}{1.05}
    \begin{tabular}{
        p{0.27\columnwidth}
        p{0.43\columnwidth}
        p{0.15\columnwidth}
        c}
        \toprule
        \textbf{Domain} & \textbf{Task Scenario} & \textbf{Difficulty} & \textbf{\#Skills} \\
        \midrule

        \multirow{4}{*}{Office \& White-Collar}
        & Citation Check                & Medium & 1 \\
        & Court Form Filling            & Easy   & 1 \\
        & Excel Table in PPT            & Medium & 2 \\
        & PDF--Excel Diff               & Medium & 2 \\

        \midrule

        \multirow{3}{*}{Software Engineering}
        & JAX Computing Basics          & Medium & 1 \\
        & LLM Prefix Cache Replay       & Medium & 2 \\
        & TicToc Unnecessary Abort Detection & Hard & 3 \\
        \midrule

        \multirow{2}{*}{Finance \& Economics}
        & Economic Detrending Correlation & Medium & 1 \\
        & SEC Financial Report          & Hard   & 2 \\
        \midrule

        \multirow{6}{*}{Natural Science}
        & Flood Risk Analysis           & Medium & 3 \\
        & Gravitational Wave Detection  & Medium & 2 \\
        & Lab Unit Harmonization        & Medium & 1 \\
        & Lake Warming Attribution      & Medium & 4 \\
        & Mars Clouds Clustering        & Hard   & 3 \\
        & Protein Expression Analysis   & Medium & 1 \\
        \midrule

        \multirow{4}{*}{\makecell[l]{Industrial \&\\ Physical Systems}}
        & Grid Dispatch Operator        & Medium & 3 \\
        & HVAC Control                  & Medium & 5 \\
        & Manufacturing-Equipment-Maintenance                  & Medium & 2 \\
        \midrule

        \multirow{2}{*}{\makecell[l]{Mathematics \&\\ Formal Reasoning}}
        & Powerlifting Coefficient Calculation & Medium & 3 \\
        & PDDL Airport Planning         & Medium & 1 \\

        \bottomrule
    \end{tabular}
\end{table}

Each selected \textsc{SkillsBench} scenario originally provides a single canonical task instance, which is insufficient for evaluating generalization beyond the observed execution. We therefore expand each scenario into a task family of 30 instances, yielding 600 instances across 20 scenarios. We first identify the functional requirements and procedural invariants of each task family, then use an LLM to generate parameterized variations in input content, numerical parameters, target entities, and execution contexts. These variations preserve the original task objectives, skill dependencies, and procedural workflows while adapting instance-specific verification criteria. We involve two human reviewers to manually review the realism of the variation dimensions. We finally generate 29 additional instances per scenario using single-slot and two-slot variations. Each generated instance undergoes automated LLM validation and manual execution verification to ensure solvability and fidelity to the original task specification.

For each scenario, the original canonical instance serves as the sole reconstruction instance, while the remaining 29 instances are held out for evaluation.  The adversary repeatedly queries the victim on the same instance until obtaining a successful trajectory, then refines its skills locally and evaluates the frozen collection on held-out instances without further victim queries or auxiliary-model assistance. Further details are provided in Appendix~\ref{app:task-generation}.

\subsubsection{Victim and Attacker Systems}
\label{subsec:systems}
We evaluate \textsc{CapaTheft} across heterogeneous victim and adversary agent systems, as summarized in Table~\ref{tab:systems}. The victim systems comprise four configurations: Claude Code powered by Claude Opus 4.8~\cite{anthropic2026opus48}, Codex powered by GPT-5.5~\cite{openai2026gpt55}, and OpenHands~\cite{wang2024openhands} powered by DeepSeek-V4-Flash-0731~\cite{deepseekai2026deepseekv4} or MiniMax M3~\cite{minimax2026m3}. All victim systems are accessed through their normal service interfaces in a black-box manner, without access to their model parameters or internal states. The adversary-controlled agents are instantiated using OpenHands with two weaker open-weight backbones: Qwen3.6-35B-A3B~\cite{qwen2026qwen36} and Gemma 4-31B~\cite{gemmateam2026gemma4}. Combining four victim systems with two adversary agents yields eight victim--adversary configurations. Throughout capability reconstruction, the adversary's backbone model, execution harness, available tools, and execution environment remain fixed, with modifications restricted to its skill collection. We use GPT-5.6-Sol~\cite{openai2026gpt56sol} as the default auxiliary model $M_{\rm aux}$ for workflow abstraction, execution analysis, differential diagnosis, and skill rewriting. To examine the sensitivity of \textsc{CapaTheft} to auxiliary-model choice, we additionally evaluate Qwen 3.8 Max~\cite{qwen38} and DeepSeek-V4-Flash-0731~\cite{deepseekai2026deepseekv4} as alternative auxiliary models. The auxiliary model is used exclusively during reconstruction and does not participate in subsequent task execution. Unless otherwise specified, we set the number of local rollouts per refinement round to $q=5$ and the maximum number of refinement rounds to $R_{\max}=5$.
\begin{table}[t]
\centering
\caption{Victim and attacker agent systems used in our evaluation.}
\label{tab:systems}
\scriptsize
\resizebox{\linewidth}{!}{
\begin{tabular}{lll}
\toprule
\textbf{Role} & \textbf{Agent Harness} & \textbf{Backbone Model} \\
\midrule
Victim & Claude Code & Claude Opus 4.8 \\
Victim & Codex & GPT-5.5 \\
Victim & OpenHands & Deepseek-V4-flash-0731 \\
Victim & OpenHands & MiniMax M3 \\
\midrule
Adversary & OpenHands & Qwen3.6-35B-A3B \\
Adversary & OpenHands & Gemma 4-31B \\

\bottomrule
\end{tabular}
}
\end{table}


\subsubsection{Baselines}
We compare \textsc{CapaTheft} against 6 representative baselines across three categories: skill-free execution, proprietary skill extraction, and trajectory-based skill learning and optimization. These baselines cover different sources and approaches for acquiring and refining skills, enabling a comprehensive comparison.

\noindent\textbf{{No Skill.}}
The adversary-controlled agent executes tasks without external skills, measuring its intrinsic task-solving capability.

\noindent\textbf{Proprietary Skill Extraction.}
This category evaluates whether directly recovering the victim's skill content or functionality enables capability transfer. It includes \emph{Query-Based Skill Extraction (QuerySE)}~\cite{wang2026skillstealing}, which obtains skill content through the victim's normal service interface. Our inspection confirms that the extracted content closely matches the victim's authentic skill instructions, and we deploy these skills on the weaker agent without modification. It also includes \emph{SigLeak}~\cite{geng2026sigleak}, which reconstructs proprietary skills by inferring their procedural content from behavioral signatures in victim execution trajectories.

\noindent\textbf{Trajectory-Based Skill Learning and Optimization.}
This category evaluates methods that derive or refine skills from execution experience. \emph{Trace2Skill}~\cite{ni2026trace2skill} is evaluated in two variants: \emph{Trace2Skill-Self (T2S-Self)}, which derives skills from the weaker agent's own execution trajectories, and \emph{Trace2Skill-Victim (T2S-Victim)}, which derives skills from successful victim executions. \emph{SkillOpt}~\cite{yang2026skillopt} optimizes skill instructions using execution feedback and validation results.

\begin{table*}[t]
\centering
\setlength{\tabcolsep}{4.5pt}
\renewcommand{\arraystretch}{1.15}
\caption{Overall capability reconstruction performance across eight victim--attacker configurations. The \emph{Victim w/ Skill} row reports each victim's $\overline{\mathrm{IPR}}$, and \textbf{bold} indicates the best attacker-side results.(\%)}
\label{tab:overall-results}
\begin{tabular}{l l cc cc cc cc}
\toprule
\multirow{2.5}{*}{\textbf{Attacker Agent}} & \multirow{2.5}{*}{\textbf{Method}}
& \multicolumn{2}{c}{\makecell{Claude Code\\ Opus 4.8}}
& \multicolumn{2}{c}{\makecell{Codex\\ GPT-5.5}}
& \multicolumn{2}{c}{\makecell{OpenHands\\ DeepSeek-V4}}
& \multicolumn{2}{c}{\makecell{OpenHands\\ MiniMax-M3}} \\
\cmidrule(lr){3-4}\cmidrule(lr){5-6}\cmidrule(lr){7-8}\cmidrule(lr){9-10}
& & $\overline{\mathrm{IPR}}$ $\uparrow$& $\overline{\mathrm{CRR}}$$\uparrow$ & $\overline{\mathrm{IPR}}$$\uparrow$ & $\overline{\mathrm{CRR}}$$\uparrow$ & $\overline{\mathrm{IPR}}$ $\uparrow$& $\overline{\mathrm{CRR}}$$\uparrow$ & $\overline{\mathrm{IPR}}$ $\uparrow$& $\overline{\mathrm{CRR}}$$\uparrow$ \\
\midrule
\rowcolor{vrow}
\multicolumn{2}{l}{\emph{Victim w/ Authentic Skill}}
& 75.82 & -- & 80.44 & -- & 78.51 & -- & 69.61 & -- \\
\midrule
\multirow{4}{*}{\makecell[l]{OpenHands\\\,\,\,\,\,\,\,\,\,\,\,\,\,\,+\\ Qwen3.6-35B-A3B}}
& No Skill    & 21.13 & --    & 21.13 & --    & 21.13 & --    & 21.13 & --    \\
& QuerySE   & 30.64 & 17.38  & 30.64 & 16.03  & 30.64 & 16.57  & 30.64 & 19.61  \\
& SigLeak & 32.13 & 20.11 & 35.22 & 23.76 & 30.18 &  15.77 & 29.04 & 16.32\\
& T2S-Self & 34.96 & 25.29  & 34.96 & 23.32  & 34.96 & 24.10  & 34.96 & 28.53 \\
& T2S-Victim & 29.12 & 14.61 & 30.44 & 15.70 & 28.83 &  13.42 & 30.06 & 18.42 \\
& SkillOpt & 27.61 & 11.85 & 28.92 & 13.13 & 27.13 & 10.45 &29.25  & 16.75\\
& \cellcolor{alrow}CapaTheft   & \cellcolor{alrow} \textbf{75.16} & \cellcolor{alrow} \textbf{98.79} & \cellcolor{alrow} \textbf{72.80} & \cellcolor{alrow}\textbf{ 87.12} & \cellcolor{alrow}\textbf{80.16} & \cellcolor{alrow}\textbf{102.88} & \cellcolor{alrow}\textbf{71.94} & \cellcolor{alrow}\textbf{104.81} \\
\midrule
\multirow{4}{*}{\makecell[l]{OpenHands\\\,\,\,\,\,\,\,\,\,\,\,\,\,\,+\\ Gemma 4-31B}}
& No Skill    & 14.94 & --    & 14.94 & --    & 14.94 & --    & 14.94 & --    \\
& QuerySE   & 35.59 & 33.92  & 35.59 & 31.52  & 35.59 & 32.48  & 35.59 & 37.77  \\
& SigLeak & 26.12 & 18.36 & 28.94 & 21.37 & 25.33 & 16.34  & 27.92 &23.74  \\
& T2S-Self & 28.81 & 22.78 & 28.81 & 21.17 & 28.81 & 21.82 & 28.81 & 25.37 \\
& T2S-Victim & 24.92 &16.39 & 24.94 & 15.27 & 25.13 &  16.03 & 24.26 & 17.05\\
& SkillOpt & 22.51 & 12.43 & 25.52 & 16.15 & 22.33 &  11.63 & 24.65 & 17.76\\
& \cellcolor{alrow}CapaTheft   & \cellcolor{alrow}\textbf{78.29} & \cellcolor{alrow}\textbf{104.06} & \cellcolor{alrow}\textbf{70.77} & \cellcolor{alrow} \textbf{85.24} & \cellcolor{alrow}\textbf{83.72} & \cellcolor{alrow}\textbf{108.20} & \cellcolor{alrow}\textbf{73.91} & \cellcolor{alrow}\textbf{107.87} \\
\bottomrule
\end{tabular}
\end{table*}
\subsubsection{Evaluation Metrics}
We evaluate capability reconstruction using two complementary metrics: \emph{Instance Pass Rate} (IPR) and \emph{Capability Recovery Rate} (CRR). For each task scenario, all methods are evaluated on the 29 held-out instances, with each instance independently executed 3 times to account for execution stochasticity. Each execution is assessed using the corresponding task-specific verifier, which returns a binary outcome indicating whether the task is successfully completed.

\noindent\textit{\textbf{Instance Pass Rate (IPR).}}
IPR measures an agent's task-solving performance on unseen instances. For scenario $s$, it is defined as
\begin{equation}
\small
\mathrm{IPR}_s
=
\frac{1}{3|\mathcal{D}_s^{\mathrm{test}}|}
\sum_{t\in\mathcal{D}_s^{\mathrm{test}}}
\sum_{j=1}^{3} z_{t,j},
\end{equation}
where $\mathcal{D}_s^{\mathrm{test}}$ denotes the held-out instances of scenario $s$, and $z_{t,j}\in\{0,1\}$ indicates whether the method successfully completes instance $t$ in its $j$-th execution.

\noindent\textit{\textbf{Capability Recovery Rate (CRR).}}
CRR measures the fraction of the victim--adversary capability gap recovered by an evaluated method. For scenario $s$, it is defined as
\begin{equation}
\small
\mathrm{CRR}_s
=
\frac{
\mathrm{IPR}_s
-
\mathrm{IPR}_s^{\mathrm{NoSkill}}
}{
\mathrm{IPR}_s^{\mathrm{Victim}}
-
\mathrm{IPR}_s^{\mathrm{NoSkill}}
},
\end{equation}
where $\mathrm{IPR}_s^{\mathrm{NoSkill}}$ and $\mathrm{IPR}_s^{\mathrm{Victim}}$ denote the performance of the corresponding adversary-controlled agent without skills and the victim with its raw skills, respectively. A CRR of 0 indicates no improvement over the skill-free adversary, while a CRR of 1 indicates complete recovery of the observed capability gap. Values exceeding 1 indicate that the adversary-controlled agent outperforms the victim.

We report macro-averaged IPR and CRR across all the evaluated task scenarios, denoted by $\overline{\mathrm{IPR}}$ and $\overline{\mathrm{CRR}}$, respectively. Higher $\overline{\mathrm{IPR}}$ indicates better task-solving performance on unseen instances, while higher $\overline{\mathrm{CRR}}$ indicates greater recovery of the victim's capability advantage.
\begin{figure*}[t]
    \centering
    \includegraphics[width=0.9\textwidth]{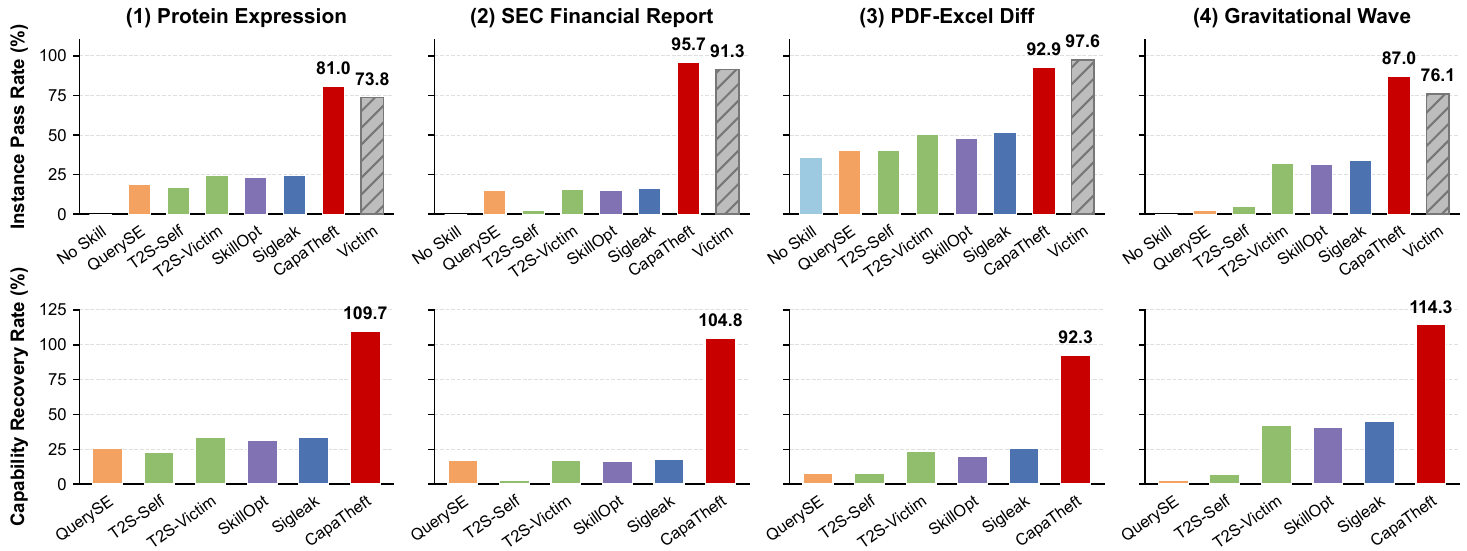}
    \caption{
    Task-level IPR and CRR of \textsc{CapaTheft} and baseline methods on 4 representative scenarios.
    }
    \label{fig:task-level-results-5}
\end{figure*}

\subsection{Attack Effectiveness}
\label{subsec:attack-effectiveness}

Table~\ref{tab:overall-results} compares the macro-averaged IPR and CRR of \textsc{CapaTheft} against representative baselines across 8 victim--adversary configurations. Figure~\ref{fig:task-level-results-5} further presents task-level results on 4 task scenarios, with the results for the remaining 16 scenarios provided in Appendix~\ref{app:ipr}.

\begin{findingbox}
\textbf{{Finding 1.}}
\textit{{CapaTheft} substantially recovers the victim--adversary capability gap.}
\end{findingbox}
Across all eight victim--adversary configurations, \textsc{CapaTheft} achieves 70.77\%--83.72\% $\overline{\mathrm{IPR}}$ and 85.24\%--108.20\% $\overline{\mathrm{CRR}}$. Compared with QuerySE, it improves $\overline{\mathrm{IPR}}$ by 35.18--49.52 percentage points. Notably, the attack agent with \textsc{CapaTheft}-curated skills even surpasses the corresponding victim in five configurations. These results demonstrate that \textsc{CapaTheft} recovers most of the victim's capability advantage through skill adaptation alone, without modifying the weaker agent's model, harness, or tools. This recovery stems from identifying missing or misexecuted behaviors through execution contrasts and translating them into fine-grained procedures adapted to the weaker agent.

\begin{findingbox}
\textbf{{Finding 2.}}
\textit{{CapaTheft} substantially outperforms existing skill extraction and optimization methods.}
\end{findingbox}
Across the eight configurations, \textsc{CapaTheft} outperforms the strongest baseline in each configuration by 35.18--48.13 percentage points in $\overline{\mathrm{IPR}}$. Existing methods primarily recover skill content or derive and refine procedural guidance from execution trajectories and feedback. However, these approaches do not explicitly identify and correct the execution deficits that prevent weaker agents from realizing the acquired procedures. Notably, directly distilling successful victim trajectories through T2S-Victim also provides limited gains, indicating that successful demonstrations alone do not ensure effective capability transfer to weaker agents. In contrast, \textsc{CapaTheft} explicitly addresses this execution gap, translating victim-derived behavioral evidence into procedural guidance adapted to the weaker agent.

\begin{findingbox}
\textbf{{Finding 3.}}
\textit{CapaTheft-driven skills generalize beyond the observed task instances.}
\end{findingbox}
For each task scenario, \textsc{CapaTheft} uses only one reconstruction instance, while evaluating the resulting frozen skills on the remaining 29 held-out instances. These instances vary in inputs, parameters, target entities, and execution contexts while preserving the original task requirements. The strong performance on held-out instances demonstrates that the \textsc{CapaTheft}-curated skills capture reusable procedural knowledge rather than memorizing the victim execution. Deployment thus requires neither further victim queries nor auxiliary-model assistance, enabling reusing the capability beyond the reconstruction instance. 

\begin{findingbox}
\textbf{{Finding 4.}}
\textit{{CapaTheft} maintains strong performance across diverse task scenarios.}
\end{findingbox}
Figure~\ref{fig:task-level-results-5} presents four representative scenarios spanning scientific analysis, financial reporting, and document processing. \textsc{CapaTheft} achieves the highest attacker-side ${\mathrm{IPR}}$ in all four scenarios, ranging from 81.0\% to 95.7\%, with corresponding ${\mathrm{CRR}}$ of 92.3\%--114.3\%. This consistency stems from its workflow-guided diagnosis of victim--adversary execution differences and adaptation of capability-critical behaviors into explicit procedures suited to the weaker agent, rather than relying on generic skill extraction or optimization. These results demonstrate its effectiveness across diverse task requirements and workflow structures.

\subsection{Transferability of CapaTheft-driven Skills}
\label{subsec:transferability}
We investigate whether skills reconstructed by \textsc{CapaTheft} can transfer across heterogeneous agent configurations without additional reconstruction. To this end, we randomly select 5 of the 20 task scenarios for transferability evaluation. Using OpenHands with Qwen3.6-35B-A3B as the source agent, we reconstruct the skill collection $\widehat{\mathcal{K}}$ for each selected scenario. The resulting skills are frozen and deployed on alternative agent configurations without further refinement or victim interaction. We consider three transfer settings: \emph{Cross-Model}, which replaces Qwen3.6-35B-A3B with Gemma 4-31B while retaining OpenHands; \emph{Cross-Harness}, which replaces OpenHands with Claude Code while retaining Qwen3.6-35B-A3B; and \emph{Cross-Model \& Harness}, which replaces both components by deploying the reconstructed skills on Claude Code with Gemma 4-31B. Table~\ref{tab:transferability} reports the macro-averaged results across the five selected scenarios, where $\Delta$ denotes the performance change relative to the source agent.
\begin{table}[t]
\centering
\setlength{\tabcolsep}{4pt}
\renewcommand{\arraystretch}{1.2}
\caption{Transferability of CapaTheft-driven skills across attacker systems, evaluated on five task scenarios (\%). }
\label{tab:transferability}
\resizebox{\columnwidth}{!}{%
\begin{tabular}{l ll ll}
\toprule
\multirow{2.5}{*}{\textbf{Transfer Setting}}
& \multicolumn{2}{c}{\textbf{Source Agent}}
& \multicolumn{2}{c}{\textbf{Target Agent} ($\Delta$)} \\
\cmidrule(lr){2-3}\cmidrule(lr){4-5}
& $\overline{\mathrm{IPR}}$ & $\overline{\mathrm{CRR}}$ & $\overline{\mathrm{IPR}}$ & $\overline{\mathrm{CRR}}$ \\
\midrule
Cross-Model
& \multirow{3}{*}{79.07} & \multirow{3}{*}{98.96}
& 72.23 \footnotesize($-$6.84) & 87.45 \footnotesize($-$11.51) \\
Cross-Harness
& &
& 84.68 \footnotesize($+$5.61) & 108.39 \footnotesize($+$9.43) \\
Cross-Model \& Harness
& &
& 83.33 \footnotesize($+$4.26) & 106.12 \footnotesize($+$7.16) \\
\bottomrule
\end{tabular}
}
\vspace{-6pt}
\end{table}
\begin{findingbox}
\textbf{{Finding 5.}}
{\textit{CapaTheft-driven skills transfer effectively across heterogeneous agent systems.}}
\end{findingbox}
As shown in Table~\ref{tab:transferability}, the reconstructed skills maintain strong performance across all three transfer settings without further adaptation. Replacing the backbone model yields 72.23\% $\overline{\mathrm{IPR}}$ and 87.45\% $\overline{\mathrm{CRR}}$. Replacing the harness further increases these metrics to 84.68\% and 108.39\%, respectively. Even when both the model and harness are replaced, the transferred skills achieve 83.33\% $\overline{\mathrm{IPR}}$ and 106.12\% $\overline{\mathrm{CRR}}$. These results demonstrate that the reconstructed skills capture reusable procedural guidance rather than execution patterns specific to the source agent. This enables a \emph{reconstruct-once, deploy-anywhere} threat: skills reconstructed on a single weak agent can be reused across heterogeneous backbones and harnesses without further victim access or skill reconstruction.
\subsection{Capability Attribution}
\label{subsec:attribution}













\begin{table}[t]
\centering
\setlength{\tabcolsep}{3pt}
\renewcommand{\arraystretch}{1.15}
\caption{Capability-Source attribution results across five scenarios(\%).}
\label{tab:source-attribution}
\footnotesize
\resizebox{\columnwidth}{!}{%
\begin{tabular}{
    p{0.28\columnwidth}
    p{0.39\columnwidth}
    l
}
\toprule
\textbf{Configuration}
& \textbf{Evidence / Auxiliary Model}
& \textbf{$\overline{\mathrm{IPR}}$ / $\overline{\mathrm{CRR}}$} \\
\midrule

\multicolumn{3}{l}{\textit{Standalone Controls}} \\

GPT-5.6-Sol
& No skills
& 28.29 / -- \\

Qwen3.6-35B-A3B
& No skills
& 7.01 / -- \\


\midrule
\multicolumn{3}{l}{\textit{Execution Evidence Ablation}} \\

Qwen3.6-35B-A3B
& Failure traces only; GPT-5.6-Sol
& 43.28/65.73 \\

Qwen3.6-35B-A3B
& Victim success + failure traces; GPT-5.6-Sol
& 71.27/116.46 \\

\midrule
\multicolumn{3}{l}{\textit{Auxiliary Model Replacement}} \\

Qwen3.6-35B-A3B
& Victim success + failure traces; DeepSeek-V4-Flash
& 70.45/114.97 \\

 Qwen3.6-35B-A3B
 & Victim success + failure traces; Qwen3.8 Max
 & 73.52/120.53 \\

\bottomrule
\end{tabular}
}
\end{table}

To isolate the sources of capability recovery, we conduct three controlled experiments on five challenging task scenarios that GPT-5.6-Sol cannot reliably solve without external skills. We use Codex powered by GPT-5.5 as the victim throughout these experiments. Specifically, we examine whether the auxiliary model’s intrinsic knowledge is sufficient for task solving, how victim execution evidence contributes to capability reconstruction, and whether the results depend on a particular auxiliary model. Table~\ref{tab:source-attribution} summarizes the results.

\begin{findingbox}
\textbf{{Finding 6.}}
\textit{Capability recovery stems from contrasting successful victim executions with weaker-agent failure traces, rather than from the intrinsic knowledge of the underlying agents or auxiliary model alone.}
\end{findingbox}

First, we examine whether the pretrained knowledge of the auxiliary model or the weaker agent alone can explain capability recovery. Without external skills, GPT-5.6-Sol and Qwen3.6-35B-A3B achieve only 28.29\% and 7.01\% $\overline{\mathrm{IPR}}$, respectively, substantially below the 71.27\% achieved by the full \textsc{CapaTheft} pipeline. This indicates that the recovered capabilities cannot be attributed to the intrinsic knowledge of either the auxiliary model or the weaker agent's backbone.

Second, we isolate the contribution of victim execution evidence by comparing two variants: one using only the weaker agent's failure trajectories and the other additionally incorporating the victim's successful reference trajectory. Adding victim execution evidence increases $\overline{\mathrm{IPR}}$ from 43.28\% to 71.27\% and $\overline{\mathrm{CRR}}$ from 65.73\% to 116.46\%. This substantial improvement shows that the weaker agent's failures alone reveal where execution breaks down but provide insufficient guidance on how to perform the missing procedures. The victim's successful execution supplies the complementary procedural reference needed to identify and reconstruct these behaviors.

Finally, we examine whether \textsc{CapaTheft} requires a strong auxiliary model by replacing GPT-5.6-Sol with two weaker alternatives, DeepSeek-V4-Flash-0731 and Qwen3.8 Max. The resulting $\overline{\mathrm{IPR}}$ values are 70.45\% and 73.52\%, respectively, compared with 71.27\% using GPT-5.6-Sol. Although performance varies across auxiliary models, both alternatives achieve substantial capability recovery. These results indicate that \textsc{CapaTheft} does not rely on a specific auxiliary model, while the model's analytical and skill-rewriting capabilities influence the attack effectiveness.

\subsection{Component Ablation}
\label{subsec:ablation}
We evaluate the three core components of \textsc{CapaTheft} on five randomly selected task scenarios, using Codex with GPT-5.5 as the victim and OpenHands with Qwen3.6-35B-A3B as the weaker agent. Specifically, \textit{w/o Workflow Abstraction} removes workflow construction; \textit{w/o Differential Diagnosis} omits victim--weaker-agent execution contrasts; and \textit{w/o Structured Skill Rewriting} replaces localized skill patching with direct rewriting. All variants retain the same reconstruction budget. Results are reported in Table~\ref{tab:ablation} in Appendix~\ref{app:ablation}.

The full \textsc{CapaTheft} achieves 79.07\% $\overline{\mathrm{IPR}}$ and 98.96\% $\overline{\mathrm{CRR}}$. Removing workflow abstraction causes the degradation, reducing $\overline{\mathrm{IPR}}$ to 49.33\% and $\overline{\mathrm{CRR}}$ to 48.93\%. This highlights the importance of extracting reusable workflow structures and cross-skill dependencies from successful victim executions. Without structured skill rewriting, $\overline{\mathrm{IPR}}$ drops to 39.20\%, demonstrating that identifying execution deficits alone is insufficient without translating them into explicit, actionable skill modifications. Removing differential diagnosis also reduces $\overline{\mathrm{IPR}}$ to 55.28\%, highlighting the importance of distinguishing capability-critical execution deficits from benign differences. These results demonstrate the complementary contributions of the three components to effective capability reconstruction.

\subsection{Practical Feasibility of CapaTheft}
\label{subsec:practical-feasibility}

\noindent\textit{\textbf{Impact of Skill Composition.}}
We evaluate \textsc{CapaTheft} on task scenarios requiring one to five coordinated skills. Across both Qwen3.6-35B-A3B and Gemma 4-31B, it maintains CRR of at least 88\% across all skill counts, without noticeable degradation as skill composition becomes more complex. These results demonstrate its effectiveness on both single-skill and multi-skill tasks. Detailed results are provided in Figure~\ref{fig:scenario-complexity} of Appendix~\ref{app:skillnum}.

\noindent\textit{\textbf{Attack Cost.}}
We measure the computational and monetary costs of capability reconstruction. On average, each task scenario requires 1.00 victim query, 6.34 local executions, and 7.34 auxiliary-model calls, consuming approximately 0.32M auxiliary-model tokens. The corresponding auxiliary-model cost is approximately \$6.32, with an average reconstruction time of 60 minutes per scenario. These results demonstrate that \textsc{CapaTheft} achieves substantial capability recovery with limited victim access and moderate reconstruction overhead, without additional model training or modifications to the weaker agent's underlying system.

\noindent\textit{\textbf{Case Study of Capability Theft.}}
We examine a business-cycle correlation task to illustrate how \textsc{CapaTheft} adapts procedural guidance to the weaker agent's execution capabilities. The victim's authentic skill specifies the high-level econometric workflow, including data cleaning, deflation, log transformation, Hodrick--Prescott (HP) filtering, and correlation analysis, but leaves certain data-parsing rules and intermediate checks implicit. For example, quarterly macroeconomic tables may omit repeated year labels in continuation rows. The victim correctly propagates the year information and validates temporal alignment, whereas the weaker agent misaligns the time series before aggregation. By contrasting their executions, \textsc{CapaTheft} identifies this deficit and patches the skill with explicit year propagation, four-quarter completeness checks, and numerical alignment verification before HP filtering. This example illustrates how implicit execution behaviors can be decomposed into concrete, verifiable procedures that weaker agents can reliably follow. The victim's authentic skill and the reconstructed skill are provided in Appendix~\ref{app:newskill}.

\subsection{Possible Countermeasures}
\label{subsec:defense-evaluation}
\begin{table}[t]
\centering
\small
\setlength{\tabcolsep}{6pt}
\renewcommand{\arraystretch}{1.25}
\caption{Effectiveness of defenses against \textsc{CapaTheft}, evaluated on 5 randomly selected task scenarios. }
\label{tab:mitigation}
\begin{tabular}{l cc}
\toprule
\textbf{Defense} & $\overline{\mathrm{IPR}}$ \textbf{(\%)} & \textbf{$\overline{\mathrm{CRR}}$ (\%)} \\
\midrule
\rowcolor{alrow}
None (undefended)          & \textbf{79.07} & \textbf{98.96} \\
\midrule
Input-Phase LLM Detector   & 72.35 & 87.65 \\
SkillGuard-5               & 74.12 & 90.63 \\
Semantic Output Filtering  & 70.84 & 85.11 \\
Trajectory Redaction       & 68.21 & 80.69 \\
\bottomrule
\end{tabular}
\end{table}
We evaluate 4 countermeasures spanning two defense paradigms in 5 randomly selected task scenarios. The first protects skill artifacts through an \emph{Input-Phase LLM Detector}~\cite{wang2026skillstealing}, which detects skill-extraction queries; \emph{SkillGuard-5}~\cite{wang2026skillstealing}, which injects defensive instructions into skills; and \emph{Semantic Output Filtering}~\cite{wang2026skillstealing}, which filters outputs potentially revealing protected skill content. The second restricts execution information exposure through \emph{Trajectory Redaction}\cite{xu2026redact}, which removes intermediate details from observable victim traces. We evaluate these defenses using Codex with GPT-5.5 as the victim and OpenHands with Qwen3.6-35B-A3B as the adversary. Table~\ref{tab:mitigation} summarizes the results.
\begin{findingbox}
\textbf{{Finding 7.}}
\textit{The evaluated  defenses mitigate but do not eliminate capability leakage, exposing a trade-off between capability protection and execution transparency.}
\end{findingbox}
The three artifact-focused defenses provide limited mitigation, with \textsc{CapaTheft} achieving 72.35\%, 74.12\%, and 70.84\% $\overline{\mathrm{IPR}}$ under the Input-Phase LLM Detector, SkillGuard-5, and Semantic Output Filtering, respectively. This is because an adversary can obtain initial skills from alternative sources, while still exploiting observable victim executions to reconstruct missing procedures. Thus, protecting skill artifacts alone does not prevent capability theft.

Trajectory Redaction provides stronger mitigation, reducing $\overline{\mathrm{IPR}}$ to 68.21\% and $\overline{\mathrm{CRR}}$ to 80.69\%. However, capability leakage persists because final deliverables, modified artifacts, and execution feedback may still expose procedural information, while the adversary retains access to its own local failure trajectories. More aggressive information suppression could further restrict leakage but may impair legitimate transparency, debugging, and execution auditability. These results suggest that effective protection requires balancing execution-information disclosure against capability confidentiality.

\section{Discussion}
\label{sec:discussion}
\noindent\textit{\textbf{Security Implications.}}
Our findings reveal a previously underexplored threat to proprietary LLM agents: reproducing their task-solving capabilities on weaker agents through legitimate service access. Unlike conventional extraction attacks targeting individual components, this threat exploits cross-agent execution asymmetry, allowing an adversary to turn limited service access into independently deployable capabilities without compromising the victim or developing a comparably powerful model. This finding broadens the scope of agent security from protecting individual components to safeguarding the end-to-end task-solving capabilities they jointly realize, highlighting a new dimension of intellectual property leakage in commercial agent services.

\noindent\textit{\textbf{Potential Mitigations.}}
Our findings indicate that protecting static skill artifacts alone cannot prevent capability theft, as normal service interactions may expose reusable procedural knowledge. Although trajectory redaction restricts access to intermediate execution details, final deliverables, modified artifacts, and execution feedback may still reveal critical procedures. A promising direction is \emph{selective and tiered disclosure}: exposing high-level execution progress and verifiable results to ordinary users while restricting fine-grained tool interactions, intermediate artifacts, and verification details to authorized auditing contexts. Such policies should balance capability protection with legitimate requirements for transparency, debugging, and execution auditability.


\section{Conclusion}
\label{sec:conclusion}

In this paper, we investigate a previously underexplored \emph{strong-to-weak capability theft} threat, where acquiring skill artifacts alone is insufficient to reproduce stronger agents' task-solving capabilities on weaker agents. We propose \textsc{CapaTheft}, which leverages victim execution evidence to identify capability-critical deficits and adapt procedural guidance to weaker agents through skill-only modifications, without changing their underlying models or execution infrastructure. Evaluations across 20 task scenarios and eight victim--adversary configurations demonstrate substantial capability recovery, generalization to unseen tasks, and cross-system transferability, highlighting the need to protect end-to-end agent capabilities beyond individual components.






%
\clearpage
\bibliographystyle{IEEEtran}
\bibliography{references}

\appendices

\section{Task Selection and Instance Construction}
\label{app:task-generation}
\subsection{Task Selection}
We select task scenarios from \textsc{SkillsBench} based on two criteria: (1) the victim successfully completes the original task using its authentic skills, whereas the weaker agent fails even when equipped with the same skills, establishing a clear \emph{skill execution gap}; and (2) the task supports variations in inputs, parameters, target objects, or application contexts without changing its original objective, required skills, workflow, tools, or verification criteria. Scenarios that do not satisfy both criteria are excluded. We ultimately select 20 scenarios across six application domains.



\subsection{Task Instance Construction and Validation}
\noindent\textit{\textbf{Instance Construction.}}
Since each \textsc{SkillsBench} scenario provides only one canonical instance, we expand it into a 30-instance task family to evaluate generalization beyond the observed execution. We first identify the invariant task requirements, including the objective, required skills, execution workflow, tool interfaces, and verification criteria, which are preserved across all instances.

We then use an LLM to propose variation slots covering task-specific content, numerical parameters, target objects, input files, and execution contexts. Each slot specifies an allowed variation dimension and its constraints. All proposed slots are manually reviewed to remove invalid, redundant, or incompatible variations. Based on the approved slots, we generate 29 additional instances using single-slot (SS) and two-slot (MS) variations. For example, document-editing tasks vary target content and formatting conditions, while cross-format reconciliation tasks vary record values, row ordering, and document representations. Variations that introduce new capability requirements, alter essential workflow stages, or require additional skills or tools are excluded. This process yields 600 instances across 20 scenarios.




\noindent\textit{\textbf{Instance Validation.}}
Each generated instance undergoes both LLM-based validation and manual verification. We examine four aspects: (1) consistency with the original task objective and capability requirements, (2) correctness of the intended variations, (3) task and environment executability, and (4) validity of task-specific verification criteria. Instances that fail either validation are revised or regenerated.

For each scenario, the original canonical instance serves as the reconstruction instance, while the remaining 29 instances are reserved for held-out evaluation. Neither held-out instances nor their execution results are accessible during reconstruction. The reconstructed skill collection is frozen before evaluation, ensuring that the reported performance reflects generalization beyond the observed instance.

\noindent\textit{\textbf{Execution and Evaluation.}}
Importantly, our construction process generates task instances rather than agent execution trajectories or evaluation outcomes. All task executions are performed by running the evaluated open- and closed-source agent systems through their respective execution harnesses and tools. We collect their actual execution trajectories and task outputs, and assess task completion using task-specific verifiers. No execution trajectories or evaluation results are synthetically generated by LLMs. This ensures that our evaluation reflects the actual execution capabilities of the tested agents rather than simulated agent behaviors.

\begin{table*}[htb]
\centering
\caption{Generalization directions for the PPTX Reference Formatting task.}
\label{tab:pptx-reference-formatting}

\small
\setlength{\tabcolsep}{5pt}
\renewcommand{\arraystretch}{1.08}
\resizebox{\linewidth}{!}{
\begin{tabularx}{\textwidth}{
    @{}
    >{\raggedright\arraybackslash}p{0.13\textwidth}
    >{\raggedright\arraybackslash}p{0.25\textwidth}
    >{\raggedright\arraybackslash}X
    @{}
}
\toprule
\textbf{Slot ID} &
\textbf{Generalization Direction} &
\textbf{Description} \\
\midrule

SS-01 &
Target Paper Identity &
Replaces target papers and synchronizes the corresponding slide headings and abstracts. \\

SS-02 &
Title Repetition Topology &
Redistributes title occurrences across slides, varying their frequencies and duplicate patterns. \\

SS-03 &
Initial Text-Box Geometry &
Changes the initial positions and dimensions of target text boxes while preserving the required final placement. \\

SS-04 &
Initial Text Formatting &
Varies fonts, sizes, colors, boldness, and run segmentation before normalization. \\

\midrule

MS-01 (SS-02 + SS-04) &
Repetition Topology and Formatting &
Combines altered duplicate patterns with varied initial formatting across target text boxes. \\

MS-02 (SS-01 + SS-04) &
Paper Identity and Formatting &
Combines new paper content with varied initial formatting while preserving the slide structure. \\

\bottomrule
\end{tabularx}
}
\end{table*}
\begin{table*}[t]
\centering
\caption{Generalization directions for the PDF--Excel Difference Detection task.}
\label{tab:pdf-excel-diff}

\small
\setlength{\tabcolsep}{5pt}
\renewcommand{\arraystretch}{1.08}
\resizebox{\linewidth}{!}{
\begin{tabularx}{\textwidth}{
    @{}
    >{\raggedright\arraybackslash}p{0.13\textwidth}
    >{\raggedright\arraybackslash}p{0.25\textwidth}
    >{\raggedright\arraybackslash}X
    @{}
}
\toprule
\textbf{Slot ID} &
\textbf{Generalization Direction} &
\textbf{Description} \\
\midrule

SS-01 &
Field-Level Numeric Change &
Changes one salary, tenure, or score value in the current workbook. \\

SS-02 &
Deleted Employee Set &
Removes one or more complete employee records from the current workbook. \\

SS-03 &
Excel Row Ordering &
Reorders complete workbook rows without changing record identities or values. \\

SS-04 &
Equivalent PDF Numeric Representation &
Reformats numeric values in the PDF while preserving their parsed values. \\

\midrule

MS-01\newline
(SS-03 + SS-01) &
Row Ordering and Field Modification &
Combines workbook row reordering with one field-level value change. \\

MS-02\newline
(SS-04 + SS-02) &
Numeric Representation and Record Deletion &
Combines equivalent PDF numeric formatting with additional deleted workbook records. \\

\bottomrule
\end{tabularx}
}
\end{table*}
\subsection{Representative Instance Examples}
To illustrate the construction process, we present two representative scenarios involving document editing and cross-format data reconciliation. Each scenario preserves its original task objective and skill requirements while varying task-specific inputs and execution conditions. Rather than listing all 30 instances, we summarize the approved variation dimensions and representative combinations. Here, \textbf{SS} denotes a single-slot variation, in which one task dimension is modified, while \textbf{MS} denotes a two-slot variation, in which multiple compatible dimensions are modified simultaneously.

\noindent\textit{\textbf{PPTX Reference Formatting.}}
This scenario requires an agent to standardize designated paper titles in a presentation and compile them into a deduplicated reference slide while preserving unrelated content. We vary presentation content and formatting conditions while retaining the original document-editing and verification requirements. Table~\ref{tab:pptx-reference-formatting} summarizes the approved variations.

\noindent\textit{\textbf{PDF--Excel Difference Detection.}}
This scenario requires an agent to compare employee records between a PDF backup and a current Excel workbook, identifying deleted records and modified fields while ignoring superficial representation differences. We vary input records and document characteristics while preserving the original comparison requirements and verification criteria. Table~\ref{tab:pdf-excel-diff} summarizes the approved variations.

\section{Example of Capability Reconstruction}
\label{app:newskill}
This appendix presents a business-cycle analysis example to illustrate how \textsc{CapaTheft} reconstructs a skill by identifying and correcting execution deficits of a weaker agent. We compare the victim's authentic skill with the CapaTheft-driven skill, highlighting the additional procedural, verification, and recovery guidance. Shaded regions indicate the modifications introduced by \textsc{CapaTheft}.

\subsection{Original Skill}
The original skill specifies the high-level econometric workflow, including data cleaning, deflation, logarithmic transformation, HP filtering, and statistical analysis. However, it leaves several execution requirements implicit, particularly year-label propagation, quarterly data completeness, and normalized price-index handling.
\begin{skill_box}{Original Skill}
\textbf{Step-by-Step Process}

1. Load and clean data: Handle missing values and ensure proper time ordering.

2. Convert to real terms: Deflate nominal values using an appropriate price index.

3. Apply log transformation to positive level variables.

4. Apply the HP filter with an appropriate $\lambda$ for the data frequency.

5. Analyze cyclical components: Compute correlations, volatilities, etc.
\end{skill_box}

\subsection{CapaTheft-driven Skill}
By contrasting the victim's successful execution with the weaker agent's failed rollouts, \textsc{CapaTheft} identifies three capability-critical deficits: incorrect year-label propagation, insufficient verification of quarterly data completeness, and improper handling of normalized price indices during deflation. It addresses these deficits by introducing explicit execution rules, intermediate checks, and recovery procedures while preserving the original econometric workflow. Specifically, the reconstructed skill requires sequential year-label propagation, verification of four-quarter completeness, and alignment of nominal values with price ratios normalized to a reference-year value of 1 before deflation. If validation fails, the agent is instructed to inspect temporal alignment and deflation calculations before recomputing the real series.
\begin{skill_box}{CapaTheft-driven Skill}

\skillupdate{
\textbf{When to use}

Use this process for business-cycle analysis involving quarterly records with omitted year labels and normalized price indices.

\textbf{Goal}

Produce aligned and validated cyclical components for statistical analysis.
}

\textbf{Procedure}

1. Load the source tables, remove non-data rows, and sort observations chronologically.

\skillupdate{
2. Propagate year labels sequentially: maintain the most recently explicit year and assign subsequent continuation rows (e.g., \texttt{II.} or \texttt{III p.}) to that year until another explicit year appears.

3. Aggregate all available quarterly observations for each year without a supplied annual value. Verify that each of the four quarters appears exactly once, with no missing or duplicated observations. Correct inconsistencies before proceeding.
}

\skillupdate{
4. Align nominal observations with the corresponding yearly price indices and apply the appropriate deflation procedure. Verify temporal alignment and the resulting real values before proceeding.
}

5. Verify that values are finite and strictly positive before logarithmic transformation. Apply the HP filter using the appropriate smoothing parameter.

6. Align cyclical components across series and compute the requested statistics.

\skillupdate{
7. Validate the final sample and results, checking year alignment, missing or non-finite values, and the validity of the computed statistics.
}

\skillupdate{
\textbf{Execution: Normalized Price-Index Deflation}

Confirm that the price index is normalized to a reference-year value of 1. Align each nominal observation with its corresponding yearly price ratio and apply ratio-based deflation. If alignment or deflation fails, check the year join, ratio interpretation, and division formula, then recompute the real series.
}

\end{skill_box}


\section{Detailed Task-Level Results}
\label{app:ipr}
\begin{figure*}[t]
    \centering

    \begin{subfigure}[t]{0.85\textwidth}
        \centering
        \includegraphics[width=0.9\linewidth]{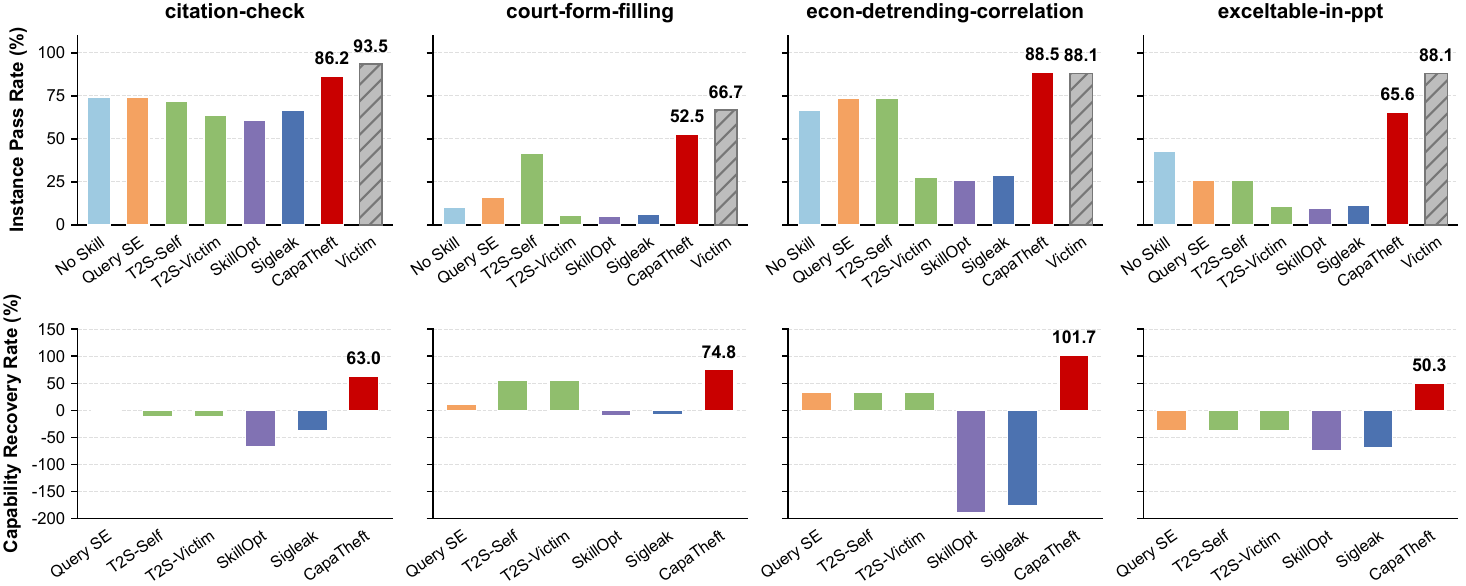}
    \end{subfigure}
    \hfill
    \begin{subfigure}[t]{0.85\textwidth}
        \centering
        \includegraphics[width=0.9\linewidth]{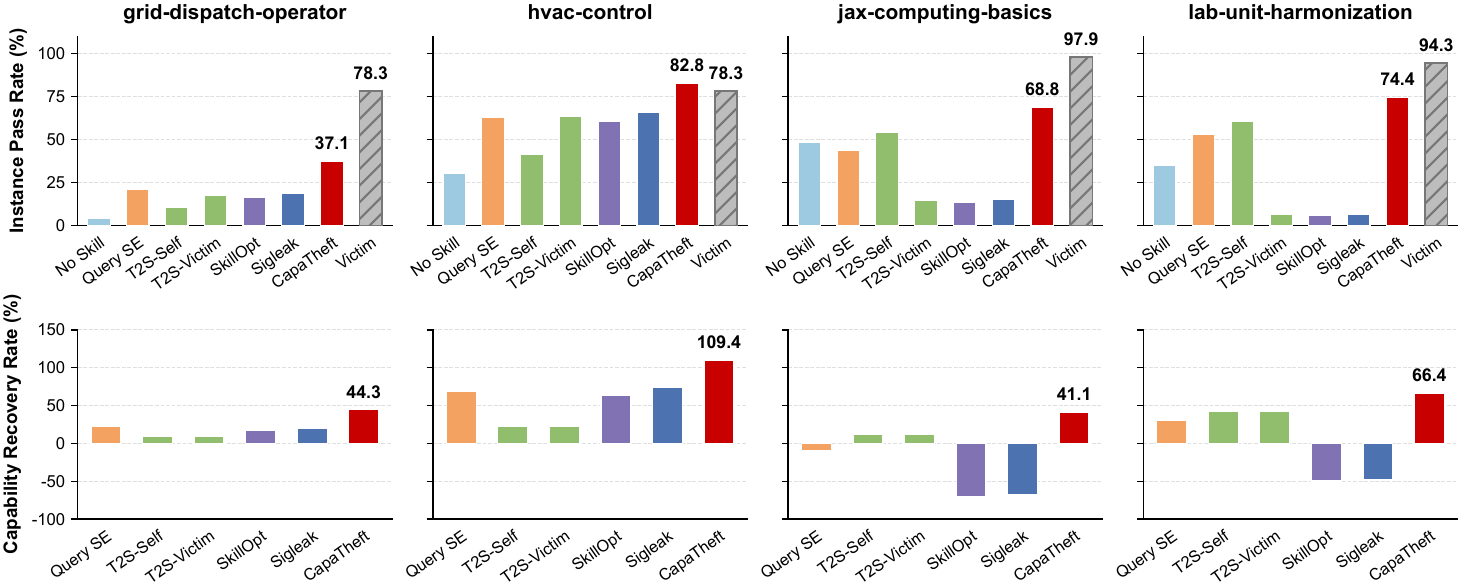}
    \end{subfigure}

    \vspace{0.6em}

    \begin{subfigure}[t]{0.85\textwidth}
        \centering
        \includegraphics[width=0.9\linewidth]{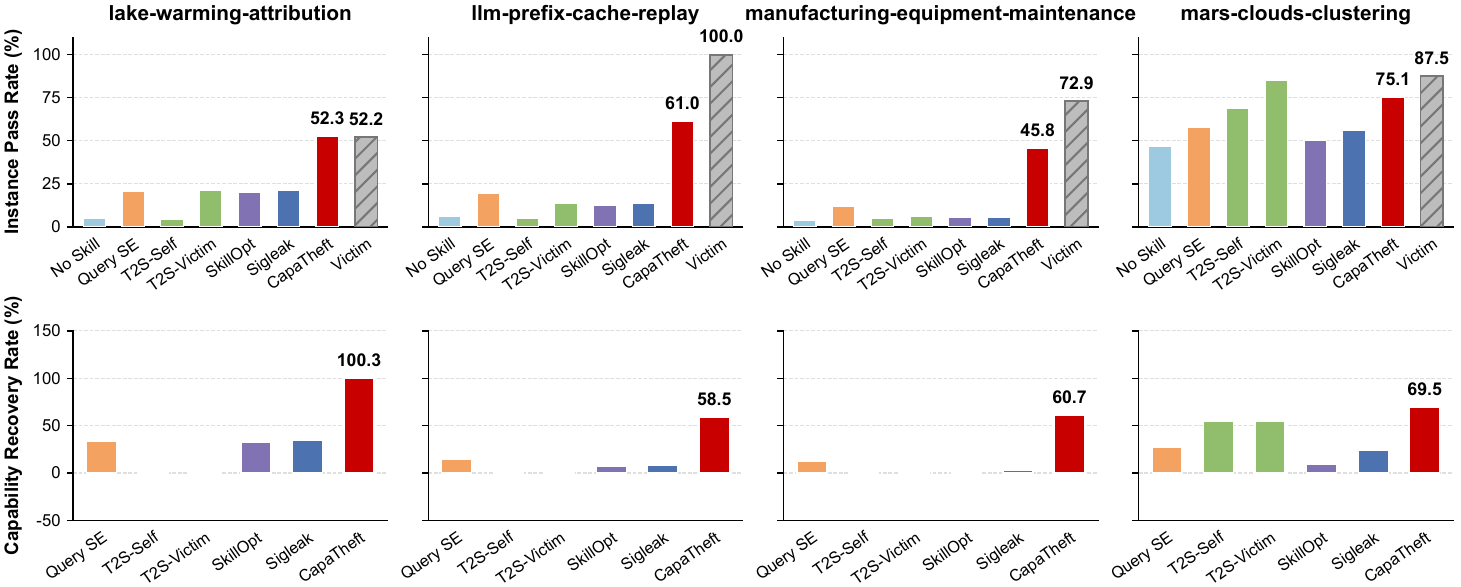}
    \end{subfigure}
    \hfill
    \begin{subfigure}[t]{0.85\textwidth}
        \centering
        \includegraphics[width=0.9\linewidth]{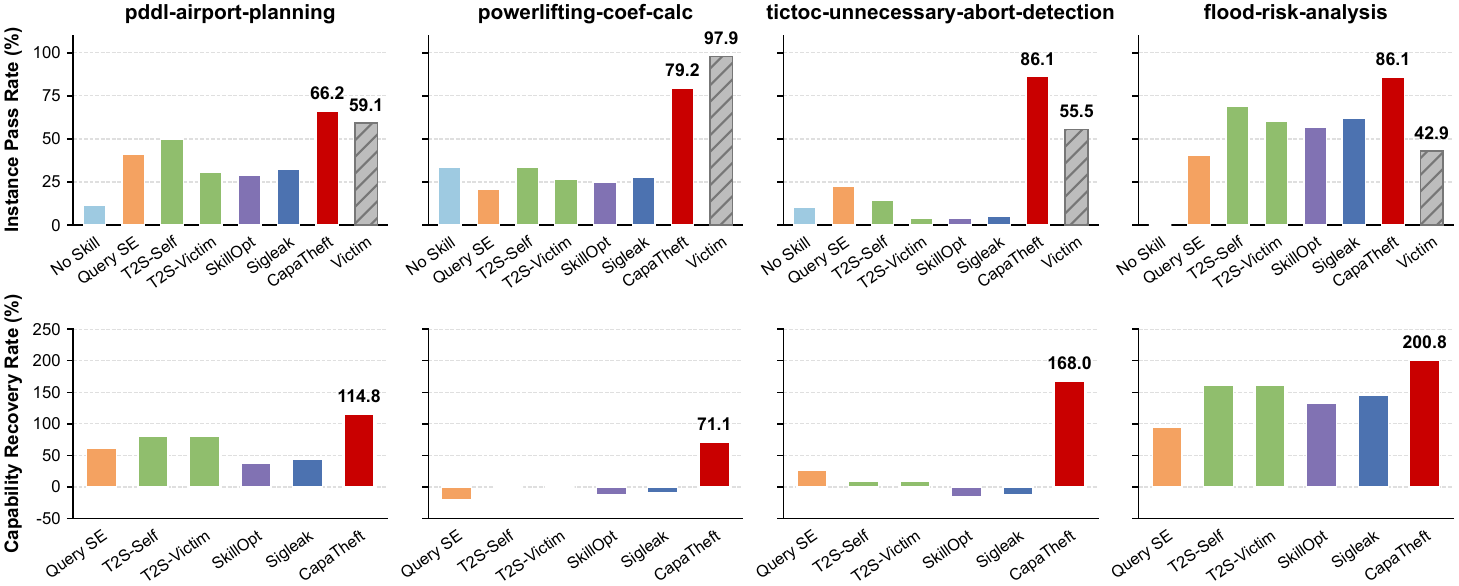}
    \end{subfigure}

\caption{Task-level IPR and CRR of the evaluated methods across the remaining 16 task scenarios.}
    \label{fig:task-level-results}
\end{figure*}

This appendix presents detailed results for the remaining 16 task scenarios, complementing the four representative scenarios reported in Section~\ref{subsec:attack-effectiveness}. Figure~\ref{fig:task-level-results} reports the scenario-level IPR and CRR across these tasks.
\begin{figure}[htb]
    \centering
    \includegraphics[width=\columnwidth]{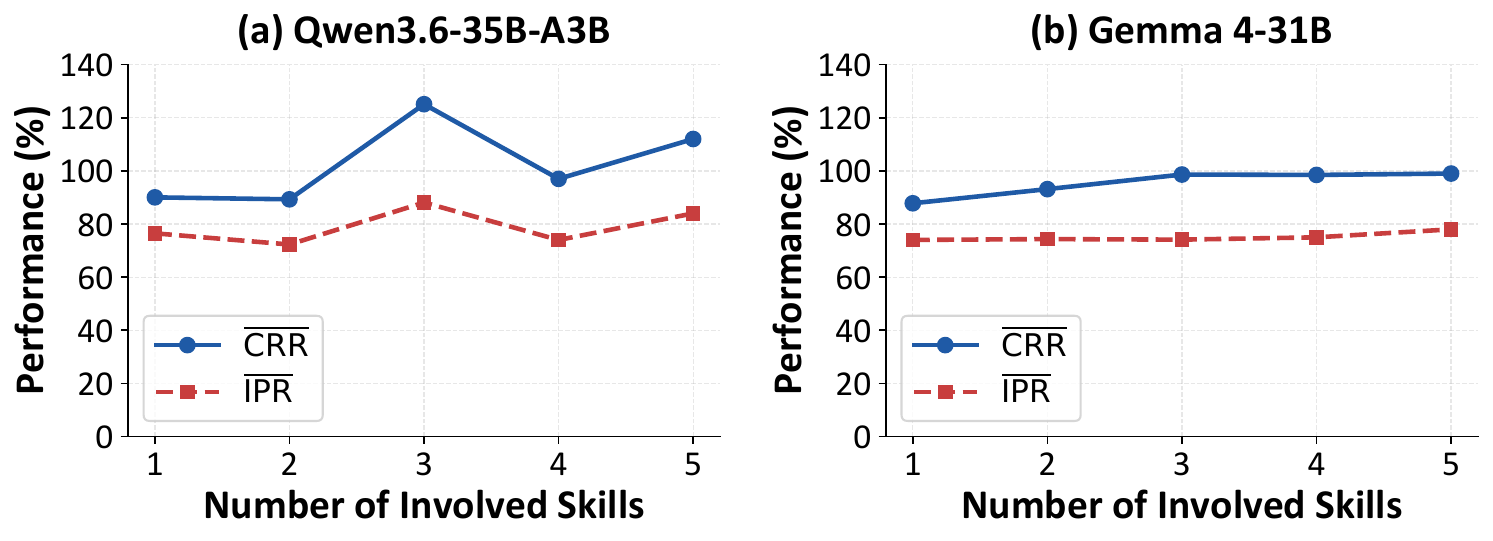}
    \caption{$\overline{\mathrm{IPR}}$ and $\overline{\mathrm{CRR}}$ of \textsc{CapaTheft} across task scenarios with different numbers of required skills.}
    \label{fig:scenario-complexity}
\end{figure}

\section{Ablation Study}
\label{app:ablation}

\begin{table}[t]
\centering
\setlength{\tabcolsep}{5pt}
\renewcommand{\arraystretch}{1.15}
\caption{Component ablation of \textsc{CapaTheft} on five randomly selected task scenarios. Parenthetical values indicate changes in percentage points relative to the full method.}
\label{tab:ablation}
\resizebox{\columnwidth}{!}{
\begin{tabular}{lcc}
\toprule
\textbf{Variant}
& \textbf{$\overline{\mathrm{IPR}}$ (\%)}
& \textbf{$\overline{\mathrm{CRR}}$ (\%)} \\
\midrule
Full \textsc{CapaTheft}& \textbf{79.07} & \textbf{98.96} \\
\midrule
w/o Workflow Abstraction & 49.33 ($-29.74$) & 48.93 ($-50.03$) \\
w/o Differential Diagnosis & 55.28 ($-23.79$) & 58.94 ($-40.02$) \\
w/o Structured Skill Rewriting & 39.20 ($-39.87$) & 31.89 ($-67.07$) \\

\bottomrule
\end{tabular}
}
\end{table}

We provide additional details of the component ablation experiments reported in Section~\ref{subsec:ablation}. We evaluate three variants, each removing one core component of \textsc{CapaTheft} while retaining the other components and the same reconstruction budget.
Table~\ref{tab:ablation} summarizes the results. Removing workflow abstraction causes the performance degradation, reducing $\overline{\mathrm{IPR}}$ from 79.07\% to 49.33\% and $\overline{\mathrm{CRR}}$ from 98.96\% to 48.93\%. This highlights the importance of extracting reusable workflow structures and cross-skill dependencies as the basis for subsequent diagnosis and reconstruction. Without structured skill rewriting, $\overline{\mathrm{IPR}}$ and $\overline{\mathrm{CRR}}$ decrease to 39.20\% and 31.89\%, respectively, indicating that identifying execution deficits alone is insufficient without translating them into actionable skill modifications. Removing differential diagnosis results in 55.28\% $\overline{\mathrm{IPR}}$ and 58.94\% $\overline{\mathrm{CRR}}$, confirming the importance of distinguishing capability-critical deficits from benign execution differences. Overall, the three components play complementary roles in reconstructing reusable task-solving capabilities.

\section{Impact of Skill Composition}
\label{app:skillnum}
Using Codex powered by GPT-5.5 as the victim, we evaluate \textsc{CapaTheft} with OpenHands powered by Qwen3.6-35B-A3B or Gemma 4-31B as the weaker agent. We group task scenarios by the number of required skills (one to five) and report their $\overline{\mathrm{IPR}}$ and $\overline{\mathrm{CRR}}$ in Figure~\ref{fig:scenario-complexity}. \textsc{CapaTheft} maintains stable performance as skill complexity increases: Qwen3.6-35B-A3B achieves above 89\% $\overline{\mathrm{CRR}}$, while Gemma 4-31B maintains 74\%--79\% $\overline{\mathrm{IPR}}$ and 88\%--100\% $\overline{\mathrm{CRR}}$.

\end{document}